%% file: main.tex
\PassOptionsToPackage{hyphens}{url}
\PassOptionsToPackage{obeyspaces}{url}
\PassOptionsToPackage{spaces}{url}
\PassOptionsToPackage{dvipsnames}{xcolor}

\documentclass[letterpaper,twocolumn,10pt]{article}

\RequirePackage[2020-02-02]{latexrelease}

\usepackage{usenix-2020-09}
\usepackage{titlesec}
\titlespacing{\section}{0pt}{6pt plus 2pt minus 2pt}{4pt plus 2pt minus 2pt}
\titlespacing{\subsection}{0pt}{6pt plus 2pt minus 2pt}{4pt plus 2pt minus 2pt}
\titlespacing{\subsubsection}{0pt}{3pt plus 2pt minus 2pt}{2pt plus 1pt minus 1pt}
\titlespacing{\paragraph}{0pt}{\parskip}{-\parskip}

\usepackage[dvipsnames]{xcolor}
\usepackage[hyphens,obeyspaces,spaces]{url}
\usepackage[utf8]{inputenc}
\usepackage{array, multirow}
\newcolumntype{P}[1]{>{\centering\arraybackslash}p{#1}}
\usepackage{tikz}
\usepackage{makecell}
\usepackage{subcaption}
\usepackage{flushend}
\usepackage{fancyhdr}
\usepackage{algorithm2e}
\usepackage{listings}
\usepackage{cprotect}
\usepackage{environ}
\usepackage{authblk}

\usepackage{pifont}
\newcommand{\cmark}{\ding{51}} %

\newcommand{\new}[1]{#1}
\newenvironment{newenv}{}{}

\usepackage{tikz}

\newcommand{\stt}[1]{{\small\path{#1}}}
\newcommand*\circled[1]{\tikz[baseline=(char.base)]{
            \node[shape=circle,draw,inner sep=1pt,semithick] (char) {#1};}}

\newcolumntype{H}{>{\setbox0=\hbox\bgroup}c<{\egroup}@{\hspace*{-\tabcolsep}}}
\newcommand{\rot}[1]{\rotatebox[origin=c]{90}{ ~#1~ }}
\newcommand{\mr}[2]{\multirow{#1}{*}{#2}}
\newcommand{\mc}[3]{\multicolumn{#1}{#2}{#3}}
\newcommand{\mytodo}[1]{}

\usepackage[printwatermark]{xwatermark}
\newwatermark[allpages,color=black!100,angle=0,scale=1,xpos=0,ypos=-131]{\small 2022 31th USENIX Security Symposium \\ Accepted version. \url{https://www.usenix.org/conference/usenixsecurity22/presentation/jeitner}}

\begin{document}

\date{}

\title{%

  XDRI Attacks - and - How to Enhance Resilience of Residential Routers
  
}

\author[*$\S$]{Philipp Jeitner}
\author[*$\S\dag$]{Haya Shulman}
\author[*]{Lucas Teichmann}
\author[*$\S\ddag$]{Michael Waidner}

\affil[$\S$]{National Research Center for Applied Cybersecurity ATHENE}
\affil[*]{Fraunhofer Institute for Secure Information Technology SIT}
\affil[$\ddag$]{Technische Universität Darmstadt}
\affil[$\dag$]{Goethe-Universität Frankfurt}

\renewcommand\Authands{ and }

\maketitle

\input{00-abstract.tex}

\input{01-introduction}

\input{02-works}

\input{03-attacks}

\input{dataset}

\input{04-1-blackboxrouters}

\input{04-2-whitebox-firmware}

\input{05-adnet-study}

\input{06-nodns}

\input{07-conclusions.tex}

    \new{
\section*{Acknowledgements}
We thank the reviewers for their helpful comments on our work. This work has been co-funded by the German Federal Ministry of Education and Research and the Hessen State Ministry for Higher Education, Research and Arts within their joint support of the National Research Center for Applied Cybersecurity ATHENE and by the Deutsche Forschungsgemeinschaft (DFG, German Research Foundation) SFB~1119.
}

{
\footnotesize
\bibliographystyle{IEEEtran}
\bibliography{main.bib,rfc2.bib,cve}
~
}

\appendix 
\input{09-appendix.tex}
\input{05-5-countermeasures}
\input{synchronization}

\end{document}

%% file: 00-abstract.tex
\begin{abstract}

We explore the security of residential routers and find a range of critical vulnerabilities. Our evaluations show that 10 out of 36 popular routers are vulnerable to injections of fake records via misinterpretation of special characters. We also find that in 15 of the 36 routers the mechanisms, that are meant to prevent cache poisoning attacks, can be circumvented.

In our Internet-wide study with an advertisement network, we identified and analyzed 976 residential routers used by web clients, out of which more than 95\% were found vulnerable to our attacks. Overall, vulnerable routers are prevalent and are distributed among 177 countries and 4830 networks.

To understand the core factors causing the vulnerabilities we perform black- and white-box analyses of the routers. We find that many problems can be attributed to incorrect assumptions on the protocols' behaviour and the Internet, misunderstanding of the standard recommendations, bugs, and simplified DNS software implementations.

We provide recommendations to mitigate our attacks. We also set up a tool to enable everyone to evaluate the security of their routers at {\small{\path{https://xdi-attack.net/}}}.

\end{abstract}

%% file: 01-introduction.tex
\section{Introduction}

\begin{figure}[t!]
    \centering
    \includegraphics[width=0.25\textwidth]{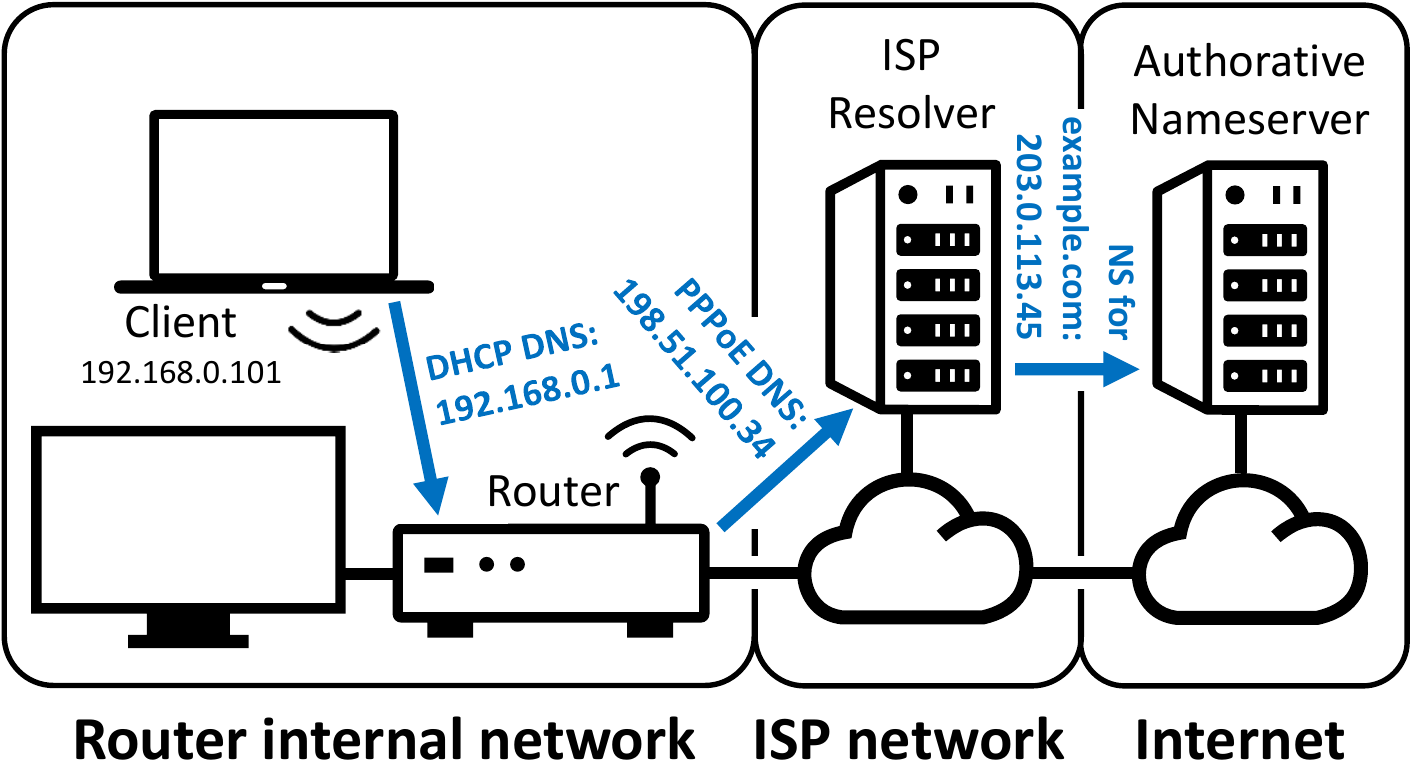}
    \caption{DNS forwarding in routers.}%
    \label{fig:dnsforwarding}
    \vspace{-12pt}
\end{figure}

Residential routers are a critical component in the Internet, connecting multiple client devices on a LAN to the network of Internet Service Provider (ISP). One of the functionalities, that residential routers provide, is forwarding Domain Name System (DNS) packets between client devices and the upstream DNS resolver, e.g., that of ISP or of a public DNS provider; we illustrate the setup in Figure \ref{fig:dnsforwarding}.
Many of the routers also provide a shared cache across the client devices on the network, aiming to improve performance. Because of their advantages, DNS forwarders in routers are widely deployed in the Internet. Studies measured that over 95\% of open DNS resolvers are forwarders \cite{schomp2013measuring}, and that a large fraction of them run on residential network devices \cite{kuhrer2014exit,nawrocki2021transparent}. Due to their central role they receive a lot of attention from the researchers and there are multiple attempts to make home routers secure and develop implementation guidelines to prevent DNS cache poisoning attacks \cite{qian:ccs20}. A DNS cache poisoning attack against a router affects all the client devices that use it, redirecting them to incorrect, potentially malicious hosts. Although such attacks are challenging to launch in practice, bugs in router implementations often allowed successful poisoning \cite{injection:router}. We describe them in Related Work, Section \ref{sc:works}. We also provide background on DNS forwarders in routers in Section \ref{ap.sc.examples} in Appendix. 

{\bf Cache poisoning residential routers with XDRI attacks.} In this work we develop practical off-path DNS cache poisoning attacks against residential routers. Our attacks use special characters to encode injections into DNS records. When those records are interpreted by a router, a misinterpretation occurs, and the router caches a poisoned DNS record. Although it was suggested that special characters could be exploited to launch Cross Site Scripting (XSS) attacks against web applications \cite{xss:dns:1,xss:dns:2,xss:dns:3,usenix-injections}, no vulnerable DNS implementations were found, hence such encodings were not considered a practical cache poisoning threat. We show for the first time how to weaponize special characters for practical DNS cache poisoning attacks against routers. The adversary only needs to control a domain name, and to create records that contain encoded injections. When the victim router receives and caches the decoded records they result in rogue mappings that redirect the client devices on that network to an adversarial server. We call our attacks {Cross-DNS Router Injections (XDRI)}. %

{\bf Attack surface of XDRI in the Internet.} We purchased 36 popular and widely deployed routers and evaluated the XDRI attacks against them. Out of 28 popular routers that implement a cache, we find 10 routers to be vulnerable to cache injections via XDRI attacks. We also evaluate the XDRI attacks against routers with open and non-open DNS forwarders, finding 8\% (100K) of the devices with open forwarders to be vulnerable and out of 973 devices with non-open forwarders, 929 to be vulnerable.

{\bf Defenses can be circumvented.} In addition to injections into caches, we test defenses against cache poisoning attacks in routers: DNSSEC [RFC4035], as well as source port and Transaction Identifier (TXID) randomization [RFC5452]. We show experimentally that although the bugs in secure selection of port and TXID were patched \cite{injection:router}, the current implementations are still vulnerable. We develop simple techniques allowing to circumvent the entropy in DNS queries, which exposes to cache poisoning attacks. We also find that an unclear standard recommendations in [RFC4035] can be exploited to disable DNSSEC validation on upstream resolvers by setting the CD bit to 1.

{\bf Root factors for vulnerabilities.} DNS forwarders in residential routers are implemented by third party software developers. The routers' developers integrate these implementations into the routers. Therefore, in order to understand the actual problems and the vulnerabilities, it is important to look inside the routers and analyze the routers' components in isolation to find the root factors causing the vulnerabilities. Previous work has not analyzed DNS inside the routers, but tested DNS implementations independently. In this work we perform a systematic analysis of the DNS forwarders' implementations in routers, including a black box evaluation of DNS cache poisoning attacks via injections as well as evaluation of our techniques for circumventing defenses against cache poisoning. We then perform white-box analysis of the firmware of the vulnerable routers and analyze the DNS forwarders implementations that we find on github. %
We provide recommendations for mitigating our attacks, however history shows that routers will likely always have bugs in DNS implementations and other attacks will be found in the future. We propose to replace the patching process with a more systematic alternative.

{\bf Unconventional proposal.} In this work we take a strategic approach at tackling the DNS vulnerabilities in routers: we propose to remove the DNS functionality from the routers entirely. We explain the primary challenges in our proposal, such as how to find the upstream recursive resolver of the ISP after failures or when clients connect to the network. We argue that the cost of our proposal, i.e., the potential increase in latency and traffic due to loss of caching on the routers, is negligible. We explain the benefits, that eliminating DNS in routers promises, including enhanced resilience and security of ISP and clients networks by reducing the threats through attacks against DNS in routers. In addition, removing DNS will simplify the routers while not requiring any changes to the clients. We show how simple configurations in the ISP DNS infrastructure would enable this.

{\bf DNS vulnerabilities in routers are not trivial to patch.} One of the significant problems with vulnerable DNS components in routers is that there is no easy way to patch them. Even when the vendors distribute a patch, this is a patch for the router but not for third party DNS software. On the other hand, when the DNS vendor updates the software, it is not clear how a client can get that update into the router.

{\bf Ethics and disclosure.} We contacted the developers of the vulnerable router vendors. Our research already resulted in 10 registered CVEs and patches.
We did ethical evaluations of the attacks reported in this work against routers in the Internet using domains that we control. This allowed us to validate the presence of the vulnerabilities without exploiting them against real victims and without causing damage to the networks nor services in the Internet. 

{\bf Contributions.} We evaluate the security of a critical component in the Internet: routers with built-in DNS forwarders.

$\triangleright$ {\em XDRI Attacks:} We develop XDRI attacks, in which we encode special characters to inject fake records into the caches of the DNS components in routers. Adversaries can use XDRI attacks to inject rogue mappings for arbitrary target domains. As a result, all the client devices are redirected to the adversarial server for services in poisoned domains. We find that 10 out of 28 routers with caches are vulnerable to our attack (the remaining 8 out of the 36 tested routers do not implement a cache).

$\triangleright$ {\em Techniques to circumvent best practices:} We develop new techniques for exposing the UDP ports and DNS TXID, circumventing the recent patches in routers. %

$\triangleright$ {\em Black-box and white-box analyses:} To understand the root factors causing the vulnerabilities we perform a black-box and a white-box analyses of the routers in our dataset. Using the results we derive the first insights into the programming bugs and architectural decisions that lead to the vulnerabilities, and isolate the responsible components in routers. %

$\triangleright$ {\em Attack surface:} We develop techniques for evaluation of our attacks against routers with open DNS resolvers as well as routers we measured via an ad network. We find that the attack surface is large: 100K devices with open resolvers and 1K routers via ad networks were found vulnerable to all our attacks. %
Our results provide just a lower bound, and the actual number of vulnerable routers is much higher.
Overall, we find vulnerable routers in 4830 autonomous systems (ASes) and in 177 countries, which shows that the vulnerabilities are prevalent, and are not specific to some region or ISP.

$\triangleright$ {\em Artifacts of our research:} We set up a tool to enable anyone to test their routers for vulnerabilities:
{\small{\path{https://xdi-attack.net/}}}. %

$\triangleright$ {\em \new{Remove DNS from} routers:} We analyze the role of the DNS component in routers and show that it is a historical artifact which is no longer essential in networks. We show that removing the DNS component from routers would improve the resilience of home networks and overall of Internet security, without incurring high latency and large traffic volumes.

{\bf Organization.} We put our work in context of related research in Section \ref{sc:works}. Our attacks are presented in Section \ref{sec:attacks}. In Section \ref{sc:dataset} we build a dataset of popular routers, evaluate our attacks against those routers and in Sections \ref{sec:blackbox} and \ref{sec:firmware-analysis} explore the root factors for vulnerabilities via black-box and white-box analysis. In Section \ref{sec:adnet-study} we present the results of our ad network study of vulnerabilities in residential routers. We recommend countermeasures to prevent our attacks, however, patching provides only a temporary solution and new attacks may be discovered in the future. Therefore in Section \ref{sc:unorthodox} we recommend to remove DNS from routers. We conclude in Section \ref{sc:conclusions}.

%% file: 02-works.tex
\section{Related Work}\label{sc:works}

 {\bf Vulnerabilities in forwarders.} In 2014 \cite{schomp2014dns} measured open resolvers and found that some of them are home routers that accept unsolicited responses without supporting best practices for DNS [RFC5452], such as validating the IP addresses, UDP ports and TXIDs in DNS responses. Accepting records from any response that they receive without validating exposes to DNS cache poisoning attacks. This is in fact a similar vulnerability to ARP poisoning via unsolicited replies \cite{bruschi2003s}. Forwarders are susceptible to fragmentation \cite{herzberg2013vulnerable,zheng2020poison} and side channel attacks \cite{qian:ccs20,man2021dns}. These attacks are not specific to routers but exploit properties in transport and IP layers, which exist only in a small fraction of servers. Recent research showed experimentally that \cite{herzberg2013vulnerable,zheng2020poison} apply to 4\% of the top-1M Alexa domains and have a cache poisoning success rate of 4\%. The applicability of \cite{qian:ccs20,man2021dns} is 11\% but these have a much lower success rates of 0.2\%. The reason for such low success rates and applicability are the properties required from the resolvers and the nameservers: \cite{herzberg2013vulnerable,zheng2020poison} must construct the spoofed fragment according to the offset and records in the second fragment in the genuine response from the nameserver, and the successful reassembly also depends on the IP ID algorithm of the nameserver, majority of which are random. The success of \cite{qian:ccs20,man2021dns} depends on the exact window size and behavior of the DNS implementation in the router, such as the global ICMP rate limit, as well as other traffic. Since these attacks need to be tailored to each specific target, they cannot be automated. In addition, all these techniques require large traffic volumes during the attack: \cite{qian:ccs20,man2021dns} require about 1M packets for scanning the ports on the target and then for hitting the correct TXID and \cite{herzberg2013vulnerable,zheng2020poison} require about 70K packets. 

 In contrast, to all the previous attack techniques against routers our XDRI attack requires sending just the encoded fake records returned by the adversarial nameserver. Our attack is also extremely effective, if it applies to the target router (i.e., the router misinterprets the special characters), then the attack is deterministic. Furthermore, our attack can be easily automated, since it does not need to be tailored to a specific operating system and DNS software implementations at the IP, transport and application layers. \new{In order to be launched our attacks do not require any a-priori information about the routers - they either apply or not. When the attacks apply, they can be executed automatically.} This allows us to test the attack for our domain against a large population of 3M routers with open forwarders and 68455 clients using an ad network. Finally, the XDRI attack requires a weaker adversary than the previous attacks. In an XDRI attack a victim client on the router network needs to visit the domain of the adversary, e.g., this can be done by placing an advertisement on an ad network pointing to objects hosted at the domain of the adversary. In contrast, the previous works require that the adversary controls a LAN machine on the victim network.

{\bf Measurements of forwarders.} A recent study \cite{nawrocki2021transparent} found that 80\% of open resolvers are transparent forwarders. These are different than the home routers that we study, since transparent forwarders do not keep cache and do not change the original IP addresses in the DNS packets they relay to the upstream resolvers. \cite{kuhrer2014exit} found that routers among other devices are exploited for amplification reflection DDoS attacks. \cite{DBLP:journals/corr/abs-2110-02224} measured prevalence of forwarders in large networks. \cite{randall2021home} found that some Customer Premises Equipment (CPE) intercept and redirect DNS queries to alternate resolvers.

 {\bf DNS cache poisoning chronicles.} Kaminsky demonstrated the first practical DNS cache poisoning attack \cite{Kaminsky08}. Since then DNS resolvers have been patched to support best practices [RFC5452] \cite{rfc5452}, including bailiwick checks \cite{rfc2181}, source port and DNS TXID randomization, and validation of their values in responses. This makes DNS resilient to off-path cache poisoning attacks. Nevertheless, different techniques for launching cache poisoning attacks were developed, including side channels to reduce the entropy in requests as well as other methodologies like injection of spoofed IP fragments into IP defragmentation caches, to bypass the randomization parameters \cite{herzberg2012security,cns:frag:dns,herzberg2013socket,herzberg2013vulnerable,brandt2018domain,stub:cache:infocom,zheng2020poison,qian:ccs20,man2021dns}. Anecdotal proposals to tunnel injections over DNS to attack web applications \cite{xss:dns:1,xss:dns:2,xss:dns:3} were not considered a threat to DNS, and in a recent study \cite{usenix-injections} considered it a theoretical threat since no DNS implementation they evaluated was found vulnerable. In this work we extend these ideas to develop injection attacks that can be exploited for DNS cache poisoning, and demonstrate practical attacks against DNS caches of popular routers.

\new{{\bf Putting our work in context.} Our novelty is in extending the DNS cache poisoning attacks to make them applicable to DNS components in routers. We develop concrete attack scenarios exploiting TXID, ports, DNSSEC and injections. The previous methods for guessing ports or TXID \cite{Kaminsky08} as well as the methods for CNAME encoding proposed in \cite{usenix-injections} all do not apply against the DNS in routers setting. In addition to these techniques, we also needed to cope with caches: in practical setup residential routers are configured to use caches of upstream resolvers. Therefore, we develop techniques to bypass the caches of upstream resolvers, to ensure that our attack packet traverses the target routers. Finally, a mere presence of a vulnerability does not imply that it can be weaponized or exploited, hence demonstrating attacks is important to evaluate the impact and significance of a problem. In addition to carrying out a systematic study of the attacks against a comprehensive set of routers, we also perform Internet-scale measurements of the vulnerabilities. Our measurements are executed with new techniques that we developed for measuring the vulnerabilities in DNS component in routers in the wild, both through ad-network and open resolvers. We found that the vulnerabilities in routers exist in multiple networks and countries. We also carried out a white-box analysis to explain the causes of the vulnerabilities. Our work provides the first analysis of the actual implementations of routers showing which programming mistakes lead to the vulnerabilities, e.g., why misinterpretations happy or why some routers use static UDP ports from the ephemeral port range. To enable router vendors as well as clients to check for vulnerabilities we develop a new test tool, which evaluates injection by misinterpretation and DNSSEC disabling attacks.}

%% file: 03-attacks.tex
\begin{figure}[t!]
    \centering
    \includegraphics[width=0.28\textwidth]{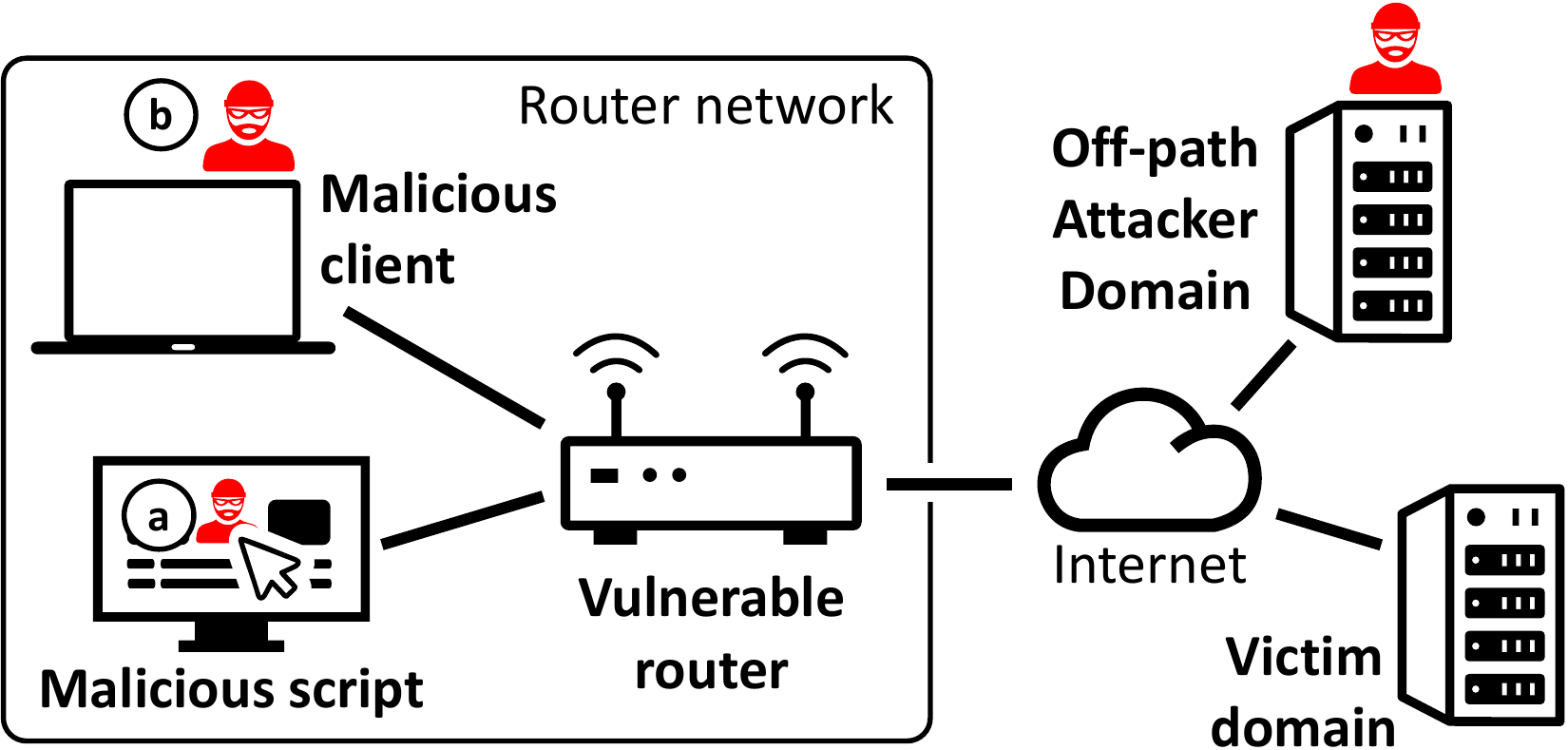}
    \caption{Adversary models.}
    \label{fig:attacker_models}
    \vspace{-12pt}
\end{figure}

\section{Cache Poisoning Attacks}\label{sec:attacks}
We developed different techniques for launching off-path DNS cache poisoning attacks. Our techniques attempt to circumvent defences in different layers of DNS in order to inject a fake record into the cache of a router. We provide overview of the attacks and the corresponding adversarial capabilities in Table~\ref{tab:attacks_overview} and in Figure \ref{fig:attacker_models}. \new{The adversary model in Figure \ref{fig:attacker_models}.\circled{b} is a superset of the model in Figure \ref{fig:attacker_models}.\circled{a}, in particular, the adversary in Figure \ref{fig:attacker_models}.\circled{a} is weaker than the adversary in \ref{fig:attacker_models}.\circled{b}.} One attack targets the records' encoding of DNS, the two other attacks enable prediction of the random challenge values and the last attack is a deactivation of DNSSEC. 

All the attacks are initiated by causing the victim to trigger DNS queries to the router. This can be done either by luring the victim to visit a website controlled by the adversary or by deploying an ad network, which causes the visiting users to issue queries. If the adversary is present on the victim network, say, the adversary controls a malware on another device, or is a legitimate user on a public network like Eduroam. We illustrate the different adversary models in Figure \ref{fig:attacker_models}.

\begin{newenv}
\textbf{Impact of DNS cache poisoning.}
It has recently been shown that DNS cache poisoning can be used to attack a variety of different applications~\cite{crosslayerattacks}. The attacks are applicable even when applications employ encryption. While many of the attacks target server-to-server protocols, which have not yet been upgraded to mandate encryption and authentication by default (such as SMTP), cache-poisoning against customer devices can still be very useful for an attacker: from shifting time on client devices via NTP (to revive expired certificates or Kerberos tickets~\cite{malhotra_attacking_2016-1}) over blocking communication with TLS-enabled websites to hijacking websites, which use un-encrypted connections. Furthermore, while our study is mainly targeted at customer-grade equipment, our results show, that our attacks also apply to some business-grade routers and in at least one case even to ISP-operated resolvers.
\end{newenv}

\begin{table}[t!]
\renewcommand{\arraystretch}{0.6}
{\footnotesize
    \centering
    \footnotesize
    \setlength{\tabcolsep}{4.5pt}
    \begin{tabular}{c|c|c|c}%
    \textbf{Section}                 & \textbf{Attack method}              & \makecell[c]{\textbf{Attacker}\\\textbf{model}}         & \textbf{Impact} \\
    \hline
    \hline
    \ref{sec:atk-injection} & \makecell[c]{Special character\\misinterpretation}  & {\small a)} ~\makecell[l]{malicious\\website}  & \makecell[c]{Cache\\poisoning} \\
    \hline
    
    \ref{sec:atk-txid}      & \makecell[c]{TXID\\forwarding}              & {\small b)} ~\makecell[l]{malicious\\client}  & \makecell[c]{Cache\\poisoning} \\
    \hline
    
    \ref{sec:atk-port}      & \makecell[c]{static\\UDP port}              & {\small a)} ~\makecell[l]{malicious\\website} & \makecell[c]{Cache\\poisoning} \\
    \hline
    
    \ref{sec:atk-cd}        & \makecell[c]{CD=1\\forwarding}              & {\small b)} ~\makecell[l]{malicious\\client}  & \makecell[c]{no DNSSEC\\protection} \\
    \end{tabular}
    \caption{Attacks overview.}
    \label{tab:attacks_overview}
    }%
    \vspace{-10pt}
\end{table}

\subsection{Cross DNS Router Injection attack}\label{sec:atk-injection}\label{sec:incomplete-cname-trigger}
Encoding attacks with special characters was proposed as a potential attack vector against applications \cite{xss:dns:1,xss:dns:2,xss:dns:3}, nevertheless \cite{usenix-injections} have not found any DNS software, appliances or devices with such vulnerabilities, hence special characters were not considered a threat for DNS or devices with DNS components, like routers. We show how to exploit special characters to develop practical attacks against routers, which we call XDRI (Cross DNS Router Injection) attack.

{\bf Adversary model.} The attack requires an adversary, that only causes a victim client to issue a query to a domain that the adversary controls. This can be done via Email or by luring a victim to visit the website of the adversary, e.g., with an advertisement network. %

{\bf Attack methodology.} Domains and hostnames are not restricted to characters. Special characters however can lead to misinterpretations. We exploit end of data and period encoding, \path{"\000"} and \path{"\."}. to develop period character and zero byte attacks, which cause the DNS software to misinterpret the domain name by manipulating the name or the subdomains.

In a zero byte attack the adversary exploits the misinterpretation of a zero byte "\path{\000}" for a string-terminator. The adversary creates a mapping between a domain that it controls, with a clever crafted subdomain \stt{www.victim.com\000.attacker.com} and an IP address that it controls.

In a period character attack the adversary exploits the misinterpretation of an period character "\path{.}" inside a DNS label for a DNS label separator. The adversary can create a record with a mapping to an IP address that it controls, say, a rec \stt{www\.victim.com.} to {\tt 6.6.6.6}. The adversary is required to control a domain under the same parent domain as the victim. When a DNS software decodes this record, it misinterprets it obtaining \stt{www.victim.com.} at \stt{6.6.6.6}.

\begin{figure}[t!]
    \centering
    \includegraphics[width=0.4\textwidth]{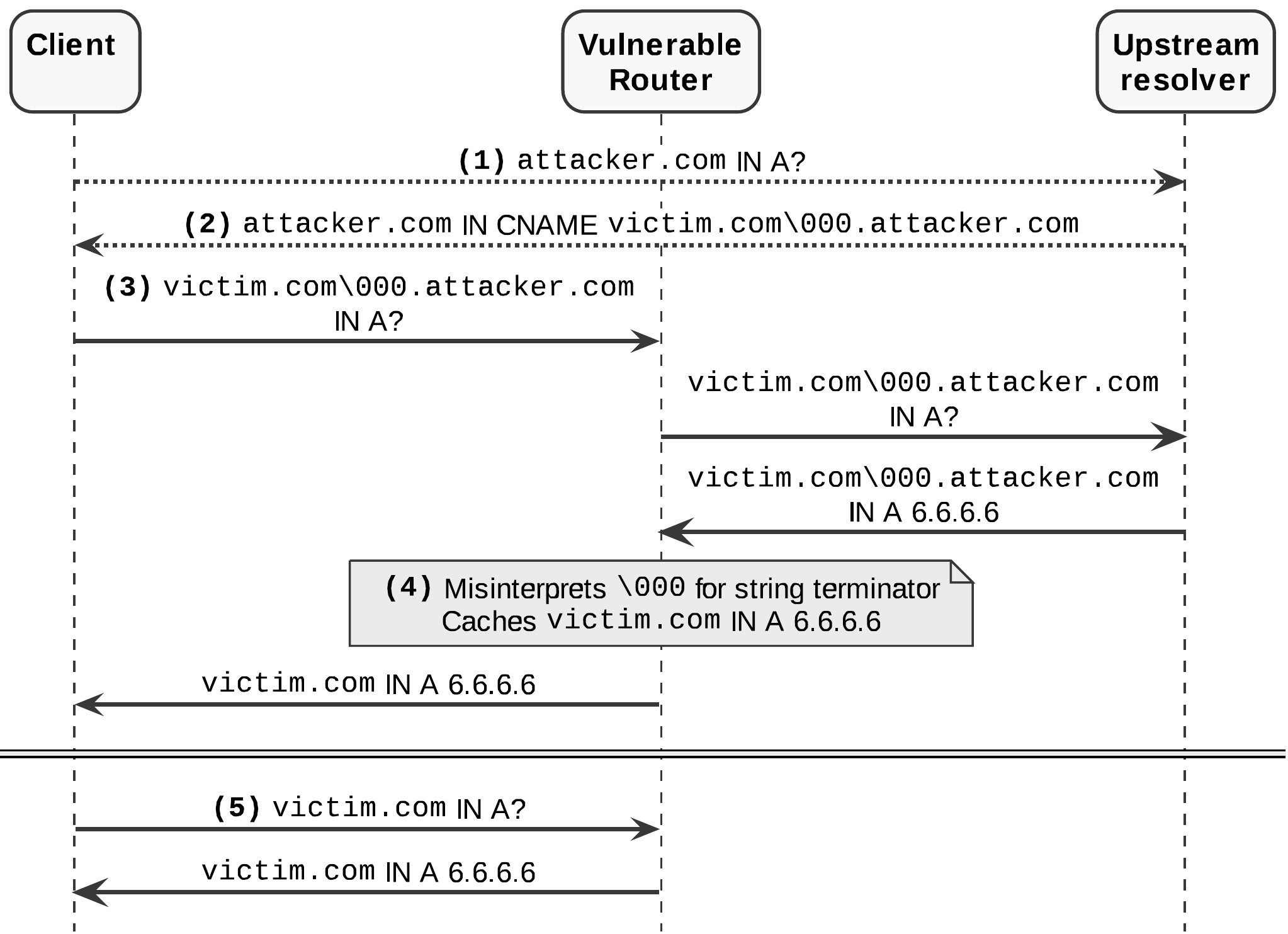}
    \caption{Cache-poisoning via character misinterpretation.}
    \label{fig:attack_injection}
    \vspace{-12pt}
\end{figure}

{\bf Attack steps.} We describe the null termination attack, illustrated in Figure \ref{fig:attack_injection}. The attack uses this CNAME record in the zonefile controlled by the adversary: \path{attacker.com} CNAME \path{victim.com\000.attacker.com}. (1) The adversary lures the victim client to visit its malicious website at \path{attacker.com}. (2) A script controlled by the adversary triggers a query to \path{victim.com\000.attacker.com}. The answer \path{victim.com\000.attacker.com} \path{IN} \path{A} \path{6.6.6.6} to this query is forwarded through the router and upstream to the attackers nameserver and the answer is provided to the router.

(4) The vulnerable router misinterprets the zero byte in the name \path{victim.com\000.attacker.com} for a string terminator, thus mapping in its cache the domain \path{victim.com} to the IP address \path{6.6.6.6} of the adversary. (5) Finally, the client visits the website at \path{victim.com} and is sent to the adversarial server at \path{6.6.6.6}.

{\bf Method to trigger queries with special characters.} Since applications, e.g., browsers, do not trigger queries with special characters, this attack is challenging to launch. In \cite{usenix-injections} the authors use a CNAME-chain to redirect a "normal" domain name, which does not contain any special characters, to a domain name which does and executed the attack by providing both records to the vulnerable forwarder in on step. In our study, we find that this attack technique does not work against most of the routers.

\begin{figure}[t!]
    \centering
    \includegraphics[width=0.45\textwidth]{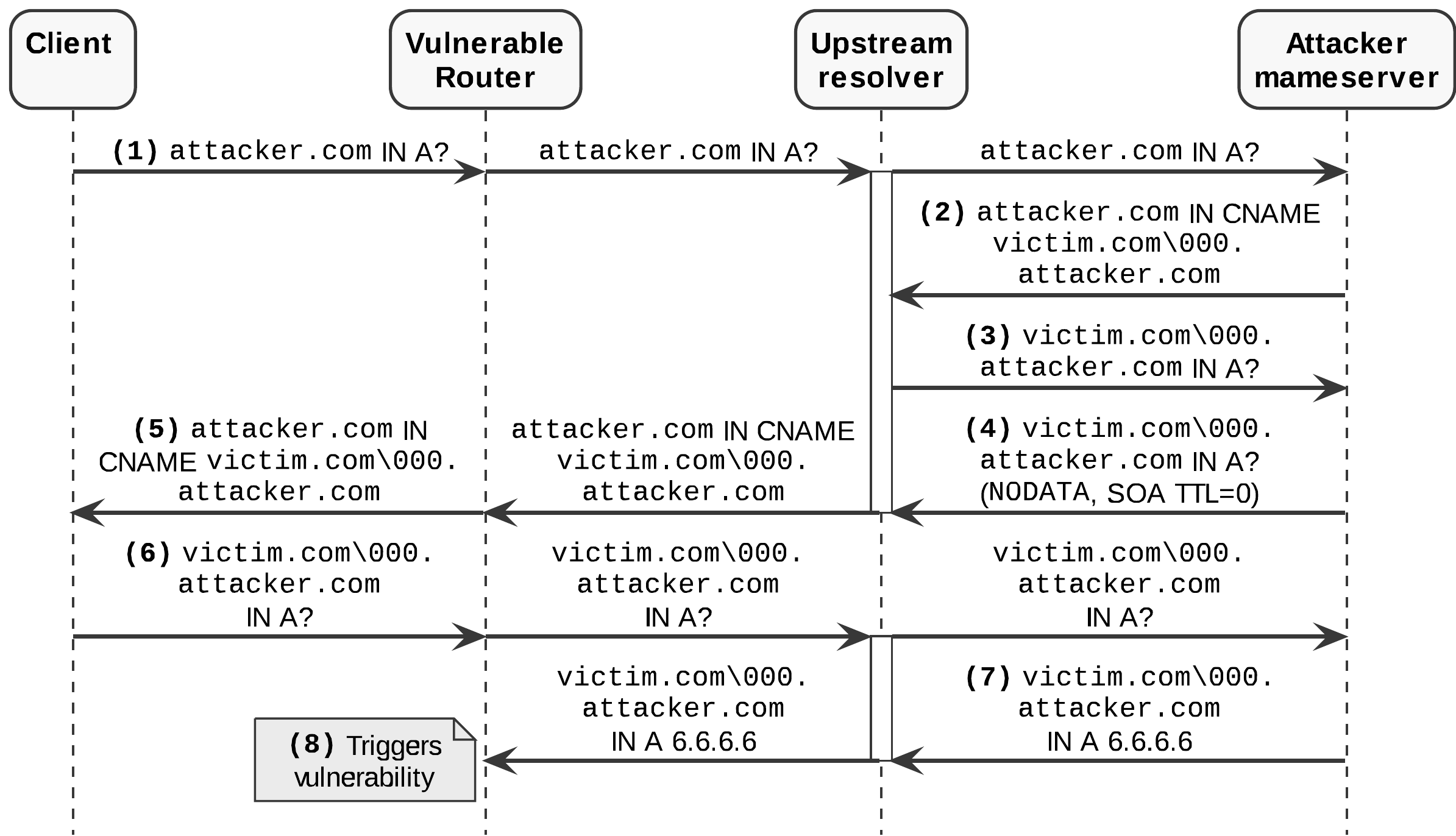}
    \caption{Special character queries with incomplete CNAME answers.}
    \label{fig:incomplete-cname}
    \vspace{-12pt}
\end{figure}

In this work we develop a method to trigger queries with special characters "\path{\000}" and "\path{.}" directly from a script in a victim client machine, even though this is disallowed by stub resolvers and browsers. Our method exploits the fact that forwarders or stub resolvers \new{used in Linux, Mac OS X and iOS} which are located downstream of the vulnerable router will chase the CNAME records upon receipt of incomplete answers and trigger queries to the CNAME target. The queries to the CNAME target are not validated - other than the queries for DNS lookups received from applications. This technique is illustrated in Figure~\ref{fig:incomplete-cname}. \new{In case the victim does not use a forwarder or stub resolver which chases CNAME records, the attack is still possible but requires the stronger attacker model from Figure~\ref{fig:attacker_models}.\circled{b} which enables the attacker to send the required queries directly to the vulnerable router.}

We prepare our custom nameserver to answer queries to \path{subdomain.attacker.com} in step (1) in Figure~\ref{fig:incomplete-cname} with a CNAME to a subdomain containing the special character \path{www.victim.com.\000.attacker.com.}. The answer however excludes the A record for this domain name (step 2). To finalize the lookup, the resolvers need to return to the nameserver and request a second query for \path{www.victim.com.\000.attacker.com.} (step 3).

However, returning an answer with a malicious encoding defeats our goal: the upstream resolver caches the misinterpreted record, and it will not be triggered by the client and traverse the router. We need the client to issue the query so that it traverses the vulnerable router. Therefore, we answer the query of the recursive resolver for \path{www.victim.com.\000.attacker.com.} with an empty answer in step 4, so that the incomplete response is forwarded to the benign client's stub resolver in step 5. Therefore in the response we include an SOA record with minimum TTL set to 0, see [RFC23080].

Once the incomplete answer reaches the stub resolver in step 6, it triggers another query for the same CNAME alias, which is this time answered by the adversarial nameserver with the malicious record mapping \path{www.victim.com.\000.attacker.com} to the IP address \path{6.6.6.6} in step 7. When this record is cached it poisons the cache of the vulnerable router because of misinterpretation (step 8).

{\bf Which resolver software chase incomplete CNAME chains?} We evaluated our technique with CNAME chasing against multiple forwarders and stub resolvers and found that the resolvers in Mac OS X, iOS and systemd-resolved chase such incomplete CNAME chains. 
As these are the default resolvers for Mac OS X, iOS and several popular Linux distributions, this technique can be used on any such system to trigger the vulnerability in routers via standard lookups, which can be triggered by any script or website visited by the user.

We list the routers vulnerable to these attacks in Table \ref{tab:allResults}: 9 routers where vulnerable against the CNAME chasing attack and only one router (Bintec) was vulnerable against the attack via CNAME.

\subsection{Reduce entropy in requests}\label{sec:atk-txid}\label{sec:atk-port}
In this section we develop simple techniques to derandomize the queries of clients behind routers. The first attack exploits the fact that some routers do not change the TXID selected by clients. We first show how to expose a TXID and then the source UDP port. The ability to predict either of the two exposes to practical DNS cache poisoning attacks. These simple attacks show that despite awareness and patches, the routers are still vulnerable. %

{\bf Adversary model.} In the attack for prediction of Transaction Identifier (TXID) the adversary is a client, e.g., a malware or a user on a wireless network. This allows the adversary to set or infer the TXID value. In the attack for inferring the source port the adversary just needs to be able to trigger queries or predict the timing when the queries are triggered, e.g., by sending an email luring the victim to visit the target website.

\begin{figure}[t!]
    \centering
    \includegraphics[width=0.4\textwidth]{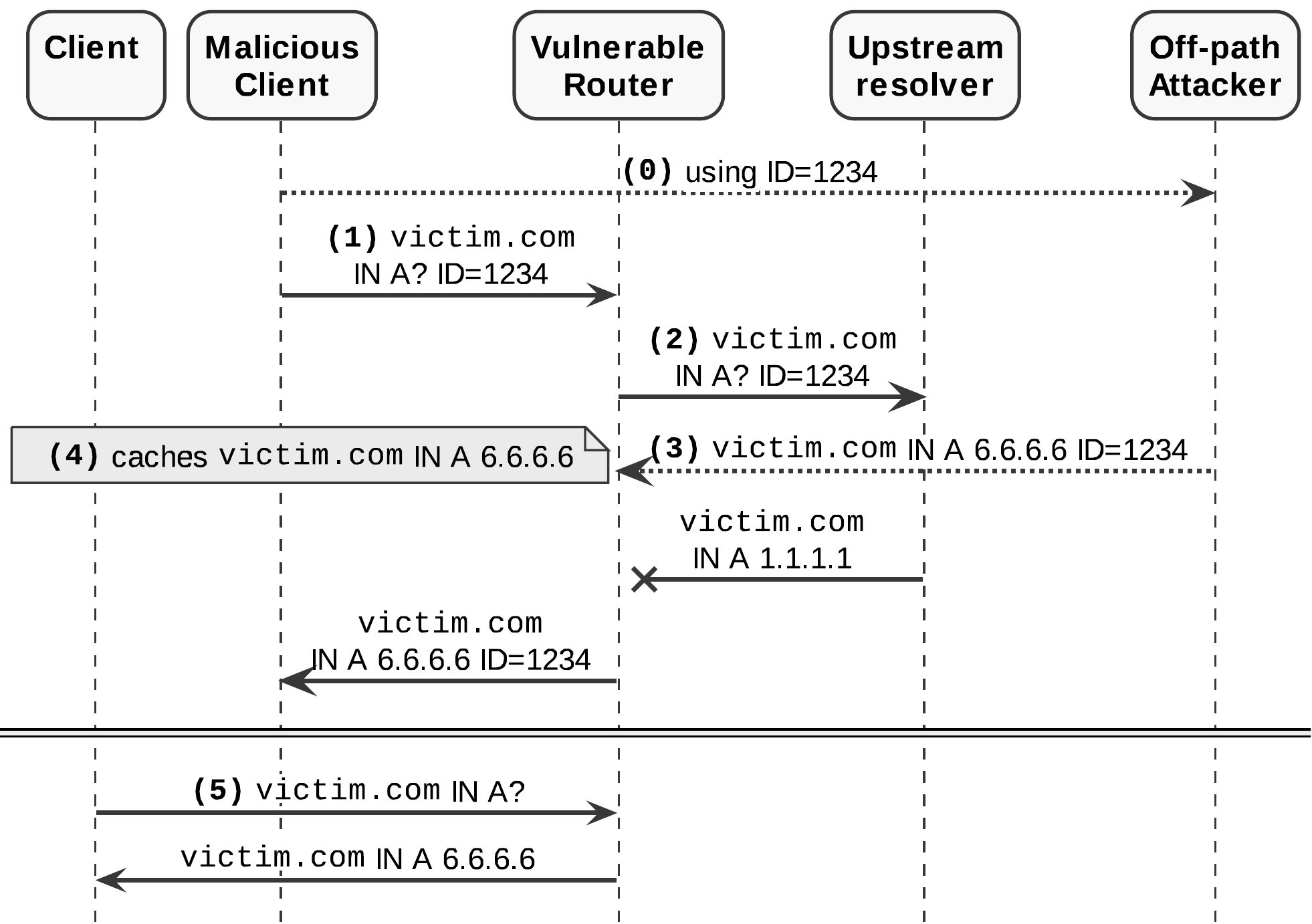}
    \caption{Off-path cache-poisoning using known TXID.}
    \label{fig:attack_txid}
    \vspace{-12pt}
\end{figure}

{\bf Recover TXID.} The steps of the attack are illustrated in Figure \ref{fig:attack_txid}. Prior to initiating the attack the malicious client communicates the DNS transaction ID it will use in the target query to the external adversary (step 0). In step 1 the adversary uses a malicious client to trigger a query to the victim domain at \path{victim.com}. The router receives the query and forwards it to the upstream resolver without changing the TXID. In step 3 the adversary can now inject a spoofed response with fake records to the vulnerable router since it already knows the TXID which is used, reducing the entropy of the challenge response parameters to 16 bits, which requires it to create at most 65536 spoofed packets. In step 4 the vulnerable router receives the injected packet and caches the address \path{6.6.6.6} for the victim domain \path{victim.com}. Finally, in step 5 the client visits the website at \path{victim.com} and is sent to the adversarial server at \path{6.6.6.6}.  

We found that 10\% of the tested popular routers do not change the TXID in queries, Table \ref{tab:allResults}.

\begin{figure}[t!]
    \centering
    \includegraphics[width=0.4\textwidth]{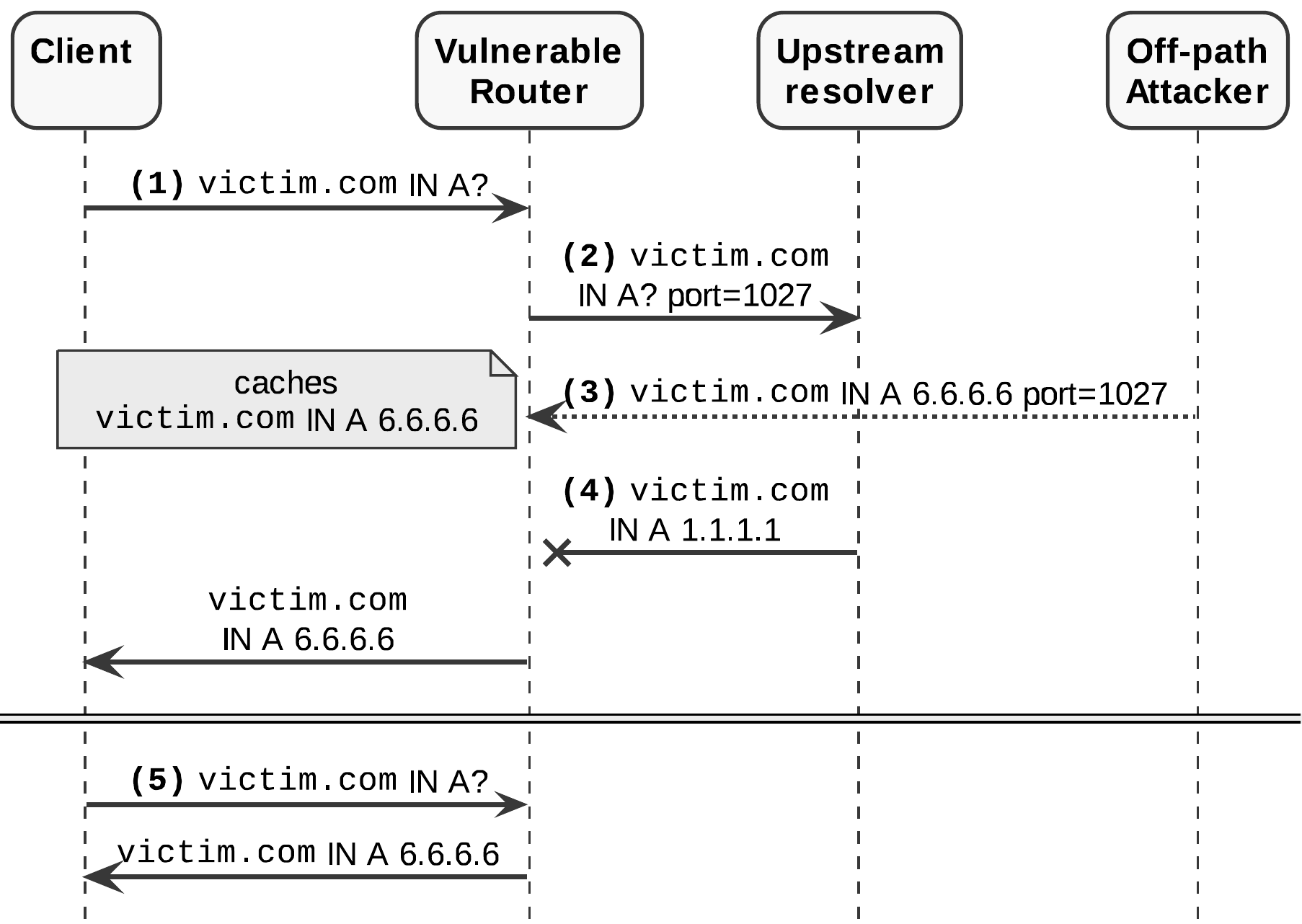}
    \caption{Off-path cache-poisoning using known UDP port.}
    \label{fig:attack_port}
    \vspace{-12pt}
\end{figure}

{\bf Recover source port.} The attack steps are illustrated in Figure \ref{fig:attack_port}. In step (1) the adversary lures the client to trigger a query to the victim website \path{victim.com}. %
The router receives the query and forwards it to the upstream resolver with a constant UDP source port in step (2). Since the adversary knows the source port used in the specific victim router implementation, it can now inject multiple spoofed responses, each with a different TXID value, until a correct response is accepted. Inferring the fixed port can be done by fingerprinting the router. We develop techniques to infer the router model using javascript in a website controlled by the adversary in Section~\ref{sec:adnet-study}. 
This attack is effective even when the victim router does not implement a cache, since the malicious answer is also forwarded to the client. We found that 19\% of the tested routers use a fixed port, Table \ref{tab:allResults}.

\subsection{Disable DNSSEC with CD flag}\label{sec:atk-cd}
The vulnerability that we describe in this section exploits a vague recommendation of the DNSSEC standard [RFC4035, Section 7], which mentions the security implications of the CD bit, but does not provide specific recommendations for caching.
Since it is not explicitly stated that resolvers should ignore the CD bit, we show that adversaries can exploit it to turn DNSSEC validation off at upstream resolvers.

\begin{figure}[t!]
    \centering
    \includegraphics[width=0.4\textwidth]{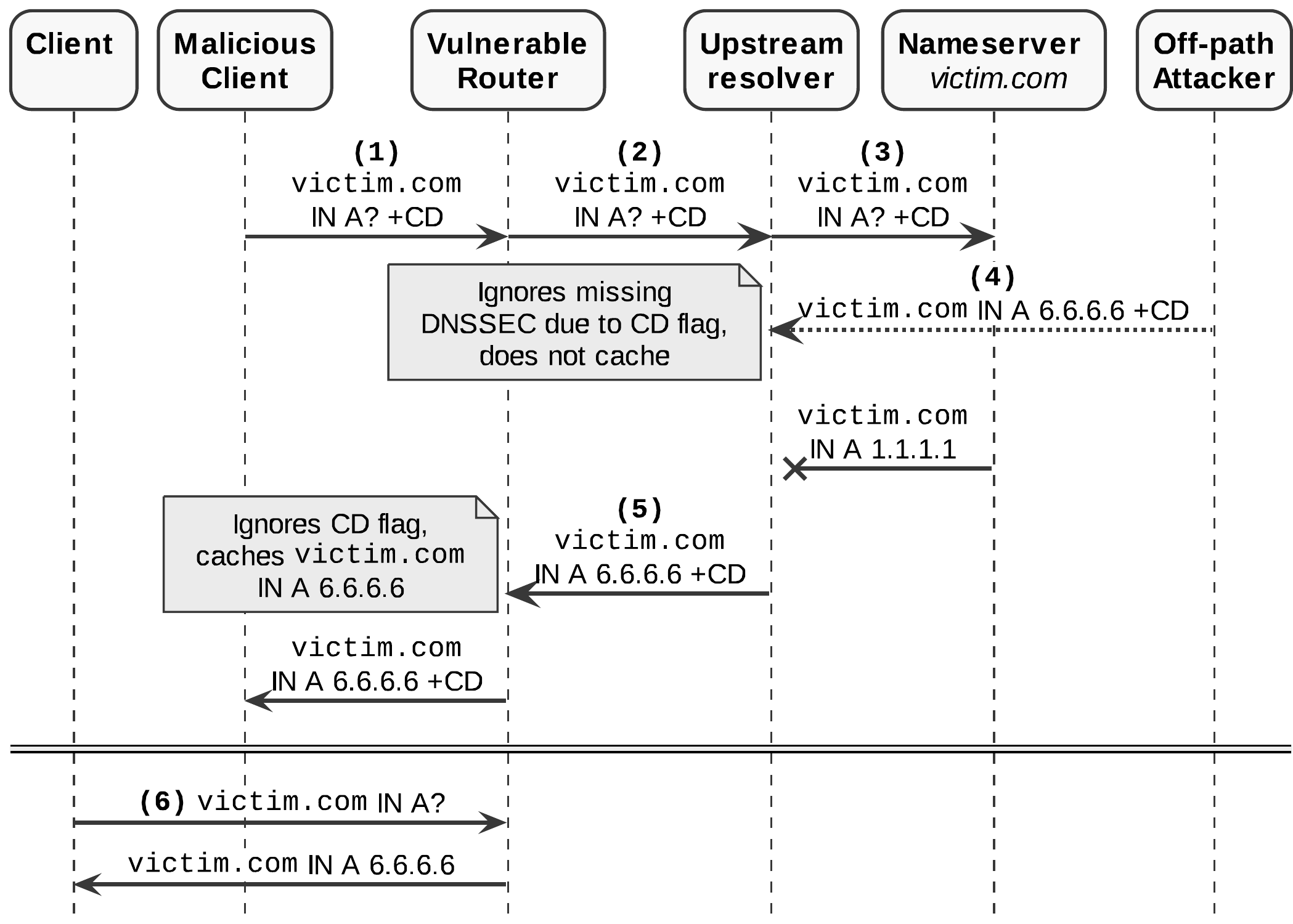}
    \caption{Disable DNSSEC with CD flag.}
    \label{fig:attack_cd}
    \vspace{-12pt}
\end{figure}

{\bf Attack steps.} The attack steps are illustrated in Figure \ref{fig:attack_cd}. First, the adversary makes a DNS query for the victim domain at \path{victim.com} to the vulnerable router using a malicious client. The query has the Checking Disabled (CD) flag set, which instructs resolvers to skip any DNSSEC checks and provide the records they receive to the client even when the DNSSEC chain of trust is broken or the DNSSEC signatures for the record do not match. In step (2) the router receives the query and forwards it to the upstream resolver without disabling the CD bit. In step (3) the upstream resolver triggers a query to the nameserver for \path{victim.com}. If the adversary can modify the records to fake values, this will not be detected since DNSSEC is not validated, but the injected record is still cached by the router since it does not acknowledge the CD flag. We explain how disabling DNSSEC can be combined with a cache poisoning attack in Section \ref{ap.sc.cache.poison} in Appendix.

\new{Applicability of our findings to other router models depends on the manufacturers’ approach in developing the firmware. Some manufacturers use the same software for all devices (i.e., AVM), while others use different software and OSes even for similar models (i.e., Tenda AC10 vs Tenda AC10v3).}

%% file: dataset.tex
\section{Dataset of Vulnerable Routers}\label{sc:dataset}\label{sec:resolverstudy}\label{sec:openresolvers}
In this section we collect the dataset of routers that we study in this work. 

{\bf Market research for popular routers.} To obtain a broad coverage of all possible router manufacturers and of different DNS forwarder implementations we select routers based on market research and with popular online shopping sites. We select manufacturers and routers from different categories: Home/SOHO (Wifi-)routers, including ISP-branded devices, small-business routers (sometimes called business/VPN routers) and Mobile/4G routers.
\new{In our router selection we tried to diversify over different manufacturers as much as possible, since this is likely to include different implementations. In our study we use all the routers we could purchase. We consult the market research sources\footnote{{\footnotesize{\href{https://www.businesswire.com/news/home/20201109005090/en/Wireless-Router-Market---Global-Industry-Analysis-Size-Share-Growth-Trends-and-Forecast-2020---2024-Technavio}{\texttt{https://www.businesswire.com/news/home/20201109005090/en/\\Wireless-Router-Market---Global-Industry-Analysis-Size-Share\\-Growth-Trends-and-Forecast-2020---2024-Technavio}}}}}$^{,}$\footnote{{\footnotesize{\href{https://www.wiseguyreports.com/reports/6365219-global-wireless-router-market-growth-status-and-outlook-2021-2026}{\texttt{https://www.wiseguyreports.com/reports/6365219-global\\-wireless-router-market-growth-status-and-outlook-2021-2026}}}}}.}

{\bf Collecting routers with open resolvers using injection attacks.} To ensure that we do not miss out on deployed devices in real ISP networks we develop a study to find the routers' types used by the ISPs. We do that in three steps. We evaluate vulnerabilities in routers with open resolvers to injection attacks via misinterpretation (Section \ref{sec:atk-injection}). In the second step we map the routers that we found to be vulnerable to their Autonomous System (AS) numbers. Third, we select ASes with high number of vulnerable devices. Our intuition is that a high fraction of vulnerable devices is likely a result of the network infrastructure of the ISP or the type of devices used by the ISPs customers, typically ISP-provided routers. For each AS where we have significant amount of data we calculate the percentage of vulnerable resolvers. We explain these steps below.

\new{The reason we use open resolvers is that in our analysis we found that many of the routers are also open forwarders, which indicated that in a dataset of open resolvers there must also be residential routers. We filter the open resolvers and concentrate the evaluation of the vulnerabilities primarily on residential routers that are configured as open forwarders. This also ensures that our evaluations do not create too much load, which would have been the result if the study was carried out by scanning the IPv4 range.}

{\em (a) Vulnerabilities to injection via special characters:} To test for the special character misinterpretation vulnerability, i.e., "\path{.}" and null terminator "\path{\000}" encoded inside DNS labels, we conduct 4 tests using different injection payloads with records in domains under our control\footnote{We use \path{test.com} and \path{target.com} as placeholders for the domains under our control.}:

{\scriptsize
\begin{verbatim}
  (1) zero.test.com CNAME www.target.com\000.test.com.
  (2) www.target.com\000.test.com. A 6.6.6.6
  (3) dot.test.com. CNAME www\.target.com.
  (4) www\.target.com. A 6.6.6.6
      www.target.com. A 1.1.1.1
\end{verbatim}
}

Each payload is designed to inject a specific domain into a router's cache. To conduct such a test, we first trigger queries for domains (1-4) where the injection payload is stored. We subsequently send a query for the domain which is to be injected by the injection payload (\path{www.target.com}) and verify that it was indeed cached. If it was cached the resolver responds with the injected IP address (\path{6.6.6.6}), otherwise the router will send an upstream query and respond with a different address (\path{1.1.1.1}).

The evaluation is performed on the 3M open resolvers from Censys \cite{censys15}. We only consider resolvers that respond to our queries for A and CNAME records  for all of the baseline tests. This results in 1.3M open resolvers from 228 different countries. 

To associate open resolvers to which we sent the query with the IP address that actually sent the query to our nameservers, we randomize all the queried domain names by prepending a random subdomain. This also allows us to avoid caching of the resolvers, just in case some components in the DNS resolution chain are shared by multiple open resolvers. To prevent other users of the resolver to be negatively affected by our tests, we run these tests only against domains we own, which allows us to validate the full injection attack without performing an attack against any other domain.

We found that 8.1\%
of the open DNS resolvers are vulnerable to injection attack. 3.8\%
of the open resolvers are vulnerable to injections with dot-misinterpretation (\path{"\."}) and 6.7\%
to injections with zero-misinterpretation (\path{"\000"}).

{\em (b) Recursive resolvers are largely not affected:} We next analyse the queries' logs on our nameservers to infer which type of resolvers are vulnerable. When we compare the IP address of the DNS query as seen by our nameserver with the IP address we originally sent the query to, we can classify the open resolvers into pure recursors (where both IP addresses match) and other types of resolvers which may forward the query through other systems. While the group of pure recursors only consists of 3.6\% (48,812 total) of all open resolvers tested, we observe that this group of resolvers is practically unaffected (0.03\% to 0.18\% vulnerable) compared to the overall results, even though it consists of open resolvers in many countries and networks. Based on this we conclude that this indicates that the misinterpretation which causes the cache-injection vulnerability must be caused by intermediate devices which implement DNS forwarding such as routers.

{\em (c) ISPs with vulnerable routers:} We found devices in 4830 Autonomous Systems (ASes) across 177 countries.
We select three ASes with the highest number of vulnerable devices: AS58265 (Level421 GmbH) with 58 out of 58 devices vulnerable, AS13787 and AS22186 (CenturyLink Communications) with 36 out of 39 vulnerable devices, and AS22394 (Cellco Partnership DBA Verizon Wireless) with 1745 out of 2105 vulnerable devices. We perform a manual research for the type of devices which are typically distributed to customers in those networks, and find that Level421 TARKAN routers on AS58265, CenturyLink Zyxel C3000z on AS13787 and AS22186, and Actiontech MI424WR on AS22394. %

Combining the routers we found through Internet study of devices vulnerable to injections with the popular routers we collected based on market research and online shops, we identify 51 different brands of routers, of which we were able to purchase 35 devices from 28 different brands\footnote{We were not able to purchase a significant number of devices, especially from smaller brands, because of import restrictions and supply shortages.}; the routers are listed in Table \ref{tab:allResults}. We use this dataset of routers in our black-box and white-box analyses. %

\new{{\bf Ethical considerations.} Our experiments were ethical. We only injected records under domains that we own (both the attack domain as well as the domain which the record is potentially misinterpreted into). Therefore, the experiment is essentially not different than merely changing the records in the zonefile on our nameservers.}

%% file: 04-1-blackboxrouters.tex
\section{Black-Box Evaluation of Vulnerabilities}\label{sec:blackbox}
In this section we develop a platform for black-box tests of routers, and evaluate our dataset of routers on that platform.
\begin{figure}[t!]
    \centering
    \includegraphics[width=0.4\textwidth]{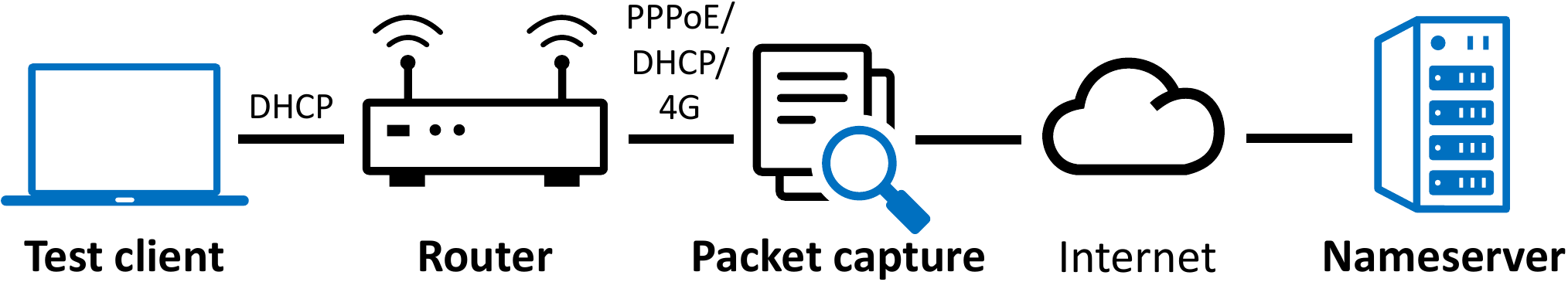}
    \caption{Black-box analysis setup.}
    \label{fig:blackbox_analysis_setup}
    \vspace{-12pt}
\end{figure}

\subsection{Platform for evaluation}
The platform we set up for testing DNS vulnerabilities in routers is shown in Figure \ref{fig:blackbox_analysis_setup}. We set up the routers for a typical configuration, changing only the settings required for an Internet connection. This includes connecting a test client to the router via its local network interface and configuring the Internet connection of a router. In most cases we configure the routers to operate in DHCP client mode, where they use an upstream network controlled by us as the internet connection. In 3 cases, where routers do not offer a DHCP client mode, we instead connect the router to the Internet via PPPoE, using a machine running OpenWRT as an PPPoE server, which emulates an ISP's broadband remote access server or (in one case) using a cellular network\footnote{In this case we capture the DNS packets at our DNS nameserver.}, as the device in question is a mobile, battery-powered hotspot without a wired interface. Additionally we change the DNS resolver used by the router to a known recursive resolver operated by us running bind version 9.16.1 either by changing its configuration or dynamically configuring it to do so over DHCP or PPPoE. 

To conduct the vulnerability tests, for each router we trigger lookups to the router's DNS forwarder using a custom-built application and observe the requests and responses emitted by the router at its Internet-side interface capturing packets as well as the responses it sends to the test client.

\newcommand{\no}{-}
\newcommand{\yes}{\cmark}

\newcommand{\da}{\no}
\newcommand{\db}{\no}
\newcommand{\dc}{\no}

\newcommand{\vinj}{(a)}
\newcommand{\vtxid}{(b)}
\newcommand{\vseq}{(c)}
\newcommand{\vport}{(d)}
\newcommand{\vcd}{(e)}

\newcommand{\vmerge}{(f)}
\newcommand{\vedns}{(g)}
\newcommand{\vntcp}{(h)}

\newcommand{\qm}{$^5$} %

\newcommand{\todo}{\textcolor{red}{\textbf{todo}}}

\begin{table*}[t!]
\renewcommand{\arraystretch}{0.8}
    \centering
    \footnotesize
    \setlength{\tabcolsep}{2.1pt} %
    \begin{tabular}{r|l|cc|c|cccc|c|c|c|c}

\multirow{4}{*}{ \textbf{Vendor} } & \multirow{4}{*}{ \textbf{Model} }                              & \multirow{4}{*}{ \makecell[c]{\textbf{Has}\\\textbf{cache}} } & \multirow{4}{*}{ \makecell[c]{\textbf{version.bind}} }      & \multicolumn{8}{c|}{\textbf{Attacks}} & \multirow{4}{*}{ \makecell[c]{\textbf{Non-}\\\textbf{standard}} } \\ \cline{5-12}

         &                                    &     & & \multirow{3}{*}{ \makecell[c]{\textbf{Any}\\\textbf{attack}} }    & \multicolumn{4}{c|}{\textbf{Misinterpretation injection}} &  \multirow{3}{*}{ \makecell[c]{\textbf{TXID}\\\textbf{forward}} } & \multirow{3}{*}{ \makecell[c]{\textbf{Fixed}\\\textbf{UDP port}} } & \multirow{3}{*}{ \makecell[c]{\textbf{CD=1 to disable}\\\textbf{DNSSEC}}} & \\
         
         &                                    &     & &      & \multicolumn{2}{c}{direct} & \multicolumn{2}{c|}{CNAME} &  &  &   & \\
         &              &   &    & & \path{\.} & \path{\000} & \path{\.} & \path{\000} &     &     &   &  \\ 
\hline
 \hline

 \multicolumn{13}{c}{ {\textbf{Home/SOHO routers}} } \\ \hline                                                                                              
 Asus        & GT-AC2900          & \yes & dnsmasq      & \no  & \no     & \no  & \no     & \no  & \db    & \dc         & \da  & \no \\ \hline                                                                                                                                                                   
 AVM         & Fritz!Box 6660     & \yes & -            & \yes & \yes    & \no  & \no     & \no  & \no    & \no         & \no & \no \\ \hline                                                                                                                                                                   
 AVM         & Fritz!Box 7312     & \yes & -            & \yes & \yes    & \no  & \no     & \no  & \no    & \no         & \no & \no \\ \hline                                                                                                                                                                   
 AVM         & Fritz!Box 7520     & \yes & -            & \yes & \yes    & \no  & \no     & \no  & \no    & \no         & \no & \no \\ \hline                                                                                                                                                                   
 AVM         & Fritz!Box 7590     & \yes & -            & \yes & \yes    & \no  & \no     & \no  & \no    & \no         & \no& \no \\ \hline                                                                                                                                                                                                            
 Cudy        & WR1300             & \yes & -$^1$        & \no  & \no     & \no  & \no     & \no  & \no    & \no         & \no  & \no \\ \hline %
 D-Link      & N150 (DIR-600)     & \yes & -$^1$        & \no  & \no     & \no  & \no     & \no  & \db    & \dc         & \da  & \no \\ \hline %
 D-Link      & N300 (DWR-920)     & \yes & -$^1$        &  \no & \no     & \no  & \no     & \no  & \db    & \dc         & \da  & \no \\ \hline %
 DrayTek     & Vigor2120          & \yes & -            & \yes & \yes\qm    & \yes\qm & \no     & \no  & \yes   & \no         & \yes& \vntcp \\ \hline                                                                                                                                                                                                                                                                                                                                                      
 Linksys     & E5350 (AC1000)     & \yes & dnsmasq-2.40 & \no  & \no     & \no  & \no     & \no  & \db    & \dc         & \da  & \no \\ \hline                                                                                                                                                                   
 Linksys     & EA8300 (AC2200)    & \yes & dnsmasq-2.78 & \no  & \no     & \no  & \no     & \no  & \db    & \dc         & \da  & \no \\ \hline                                                                                                                                                                   
 Netis       & AC1200             & \yes & dnsmasq-2.79 & \no  & \no     & \no  & \no     & \no  & \db    & \dc         & \da  & \no \\ \hline                                                                                                                                                                                                                                                                                                                                                                                                        
 Tenda       & AC10v3             & \yes & -            & \yes & \yes    & \yes & \no     & \no  & \no    & \yes (62066)& \yes& \vmerge,\vntcp \\ \hline %
 Tenda       & F3                 & \yes & -            & \yes & \yes    & \yes & \no     & \no  & \no    & \yes (50387)& \yes& \vmerge,\vntcp \\ \hline %
 Trendnet    & TW100-S4W1CA       & \yes & -            & \yes & \no     & \no  & \no     & \no  & \no    & \yes (5530) & \yes & \vmerge,\vntcp \\ \hline %
 Xiaomi      & MiRouter4A         & \yes & dnsmasq-2.71 & \no  & \no     & \no  & \no     & \no  & \db    & \dc         & \da  & \no \\ \hline                                                                                                                                                                   
 Zyxel       & Speedlink 5501     & \yes & dnsmasq-2.57 & \no  & \no     & \no  & \no     & \no  & \db    & \dc         & \da  & \no \\ \hline                                                         
 Edimax      & N300               & \no  & -            & \yes & \no     & \no  & \no     & \no  & \no    & \yes (1027) & \no  & \vntcp \\ \hline %
  Mercusys    & MW305R             & \no  & -            & \yes$^2$ & \no     & \no  & \no     & \no  & no$^2$ & \no         & \no  & \vntcp \\ \hline %
 Netgear     & AC1200 / R6120     & \no  & -            & \no  & \no     & \no  & \no     & \no  & \no    & \no         & \no  & \no \\ \hline %
  STRONG      & Wi-Fi Router 300   & \no  & -            & \yes & \no     & \no  & \no     & \no  & \no    & \yes (1027) & \no  & \vntcp \\ \hline %
 TP-Link     & Archer C7 (AC1750) & \no  & -            & \no  & \no     & \no  & \no     & \no  & \no    & \no         & \no  & \no \\ \hline %
 TP-Link     & TL-WR841N          & \no  & -            & \no  & \no     & \no  & \no     & \no  & \no    & \no         & \no  & \no \\ \hline %
 \hline
 
 \multicolumn{13}{c}{ {\textbf{ISP-branded home routers}} } \\ \hline      
 CenturyLink & C3000Z             & \yes & -            & \yes & \yes    & \yes & \no     & \no  & \yes   & \yes$^3$    & \yes & \vedns,\vntcp \\ \hline %
 Telekom     & Speedport Smart 3  & \yes & -$^1$        & \no  & \no     & \no  & \no     & \no  & \no    & \no         & \no  & \no \\ \hline %
 Vodafone    & Station TG3442DE   & \yes & dnsmasq-2.78 & \no  & \no     & \no  & \no     & \no  & \db    & \dc         & \da  & \no \\ \hline    
 Actiontec   & MI424WR            & \no  & -            & \yes & \no     & \no  & \no     & \no  & no$^2$ & \yes (1024) & \no  & \vntcp \\ \hline %
                                                                       \hline  
 \multicolumn{13}{c}{ {\textbf{Small business routers}} } \\ \hline
 Bintec      & RS353a             & \yes & -            & \yes & \no   & \no & \yes    & \yes & \no    & \no         & \no & \vmerge,\vntcp \\ \hline %
 Cisco       & RV260              & \yes & dnsmasq-2.78 & \no  & \no     & \no  & \no     & \no  & \db    & \dc         & \da  & \no \\ \hline                                
 Grandstream & GWN7000            & \yes & dnsmasq-2.78 & \no  & \no     & \no  & \no     & \no  & \db    & \dc         & \da  & \no \\ \hline   
 Synology    & RT2600AC           & \yes & dnsmasq-2.78 & \no  & \no     & \no  & \no     & \no  & \db    & \dc         & \da  & \no \\ \hline      
 Ubiquiti    & EdgeRouter4        & \yes & dnsmasq-2.78 & \no  & \no     & \no  & \no     & \no  & \db    & \dc         & \da  & \no \\ \hline      
 
 Lancom      & 1640e              & \no  & -            & \no  & \no     & \no  & \no     & \no  & \db    & \dc         & \da  & \vntcp \\ \hline  
 
 \hline

 \multicolumn{13}{c}{ {\textbf{Mobile/4G routers}} } \\ \hline                                                                                                                                    
 Huawei      & 5G CPE 5 Pro 2     & \yes & -            & \yes & \no     & \yes\qm & \no     & \no  & \no    & \no         & \no& \vedns \\ \hline                                                                                                                                                                   
 Level421    & TARKAN             & \yes & dnsmasq-2.51 &\no$^4$&\no$^4$ & \no  & \no$^4$ & \no  & \db    & \dc         & \da  & \no \\ \hline    
 Teltonika   & RUT950U022C0       & \yes & dnsmasq-2.81 & \no  & \no     & \no  & \no     & \no  & \db    & \dc         & \da  & \no \\ \hline         
 \hline
                                                                                                                                       
\multirow{2}{*}{SUM(\yes)}
 & 36     & 28 & - & 15  & 8     & 5  & 1    & 1  & 4    & 7         & 5  & 11 \\ 
 & 100\%     & 78\% & - & 42\%  & 22\%     & 14\%  & 3\%    & 13\% & 11\%    & 19\%         & 14\% & 31\% \\ \hline     

\end{tabular}\\
{\yes: vulnerable/yes. \no: not vulnerable/no. $^1$: hidden dnsmasq version, $^2$: uses sequential TXIDs, $^3$: Port selected randomly at boot. $^4$: ISP network vulnerable. \\ \qm: Query section mismatch. \vmerge: CNAME chain merging. \vedns: EDNS can cause broken responses. \vntcp: No TCP support.}
    \caption{\centering Results of the black-box analysis.}%
    \label{tab:allResults}
    \vspace{-12pt}
\end{table*}

\subsection{Evaluation methodology}

We develop methods for evaluation of vulnerabilities to injection via misinterpretation attacks as well as to TXID forwarding, UDP port prediction and disabling DNSSEC. \new{We send 5 queries per test to identify static values. We use other manual tests to verify the results of firmware analysis in Section \ref{sec:firmware-analysis}.}

\textbf{Special character misinterpretation.} The evaluation follows the same study methodology as described in Section \ref{sc:dataset}, with the only difference that the routers are set up on the premise of our network, as described in Figure \ref{fig:blackbox_analysis_setup}.

\textbf{TXID forwarding.} To test the router for TXID forwarding, i.e., whether it re-uses the TXID from client requests for its upstream packets, we send the router's DNS forwarder a few queries for different domains and compare the TXIDs from the client requests to the ones in the upstream queries. If the TXIDs are equal, we call a router vulnerable to this vulnerability if it implements a cache, otherwise the router may re-use the TXIDs, but an adversary may not be able to exploit this, because there is no cache to poison at the router in question. If the TXIDs were not equal, we do also identify if the TXIDs are sequentially allocated by manual inspection which does also represent a vulnerability.

\textbf{Static UDP port.} To test if a router uses a static UDP source port, we send the router some queries for different domains, and observe the UDP source port in the requests sent to the upstream resolver. When a router re-uses the same port multiple times for different queries, we call the router vulnerable. Furthermore, in this test we differentiate between semi-random random ports which are allocated once when the forwarder is started and fully static port numbers. To test this, after we identified that a router re-uses port numbers, we restart the router by disconnecting its power supply, and see if the port number has changed after the reboot. For this vulnerability, we do consider a router vulnerable even if it does not implement a cache, as an attacker might still inject malicious responses if he manages to conduct the attack at the point in time when a client triggers a request for the target domain.

\textbf{CD=1 forwarding.} To test DNS forwarder for disabling DNSSEC, we conduct the following test. (1) We send a query for an improperly DNSSEC-signed domain with CD bit set. We check if the query is forwarded using the CD bit set, and if the client receives an answer to its query with an A record included. (2) We then send another query for the same domain, but without the CD bit set. When the forwarder answers the query from its cache with the IP address from the improperly-signed DNSSEC response, the router is vulnerable to DNSSEC disabling attacks. 

\textbf{Fingerprint of routers.} We aim to identify the DNS forwarder implementation used in the router using \path{version.bind} queries. These special DNS queries for the domain \path{version.bind} with type \path{TXT} and class \path{CHAOS} are implemented by many DNS servers to communicate the server implementation and version. 

\textbf{TCP support.} We test DNS forwarders for DNS over TCP support, which is mandatory in [RFC5966]. We also attempt to verify if the forwarder implements a cache at all, by sending it a few queries for the same (domain,type)-tuple and see if it emits multiple queries to its upstream resolver. We record any other non-standard behaviour of the routers DNS forwarder during our tests, such as the inability to properly handle (or ignore) EDNS OPT records.

\subsection{Results}

Out of the 36 routers tested, all routers implemented some form of DNS forwarding service.  8 of the routers do not implement a cache, making them naturally invulnerable to most of the attacks. %

\textbf{Attacks.} We find 10 routers to be vulnerable against the misinterpretation attack, out of which 6 where vulnerable to the attack variant using zero-byte (\path{\000}), which allows adversaries to inject fake records for arbitrary domains.
2 routers were vulnerable to the TXID forwarding attack and another 2 used sequential TXIDs, which also makes them vulnerable. 7 routers (from 6 different vendors) did not use randomized UDP source port for upstream queries making them vulnerable to off-path injection attacks. Finally, 5 routers from 4 vendors were vulnerable to attack disabling DNSSEC protection against upstream resolvers. In total we found 15 routers from 10 different vendors vulnerable to any of our analyzed vulnerabilities, which makes 42\% of the tested devices.

 One router, the Level421 Tarkan which is a 4G mobile hotspot sold by an ISP providing special, world-wide roaming services was only found vulnerable to the misinterpretations injection attack when using the Internet connection provided by the vendor via the hotspot's embedded SIM card, but not when using the Internet connection via a customer-provided SIM card. Since the hotspot uses dnsmasq as its forwarder implementation as discovered by \path{version.bind} queries, we conclude that the vulnerability discovered in this test is caused by resolvers in the provider's infrastructure, not the actual device provided to the user.
 
\textbf{Standard compliance.} Additionally to the vulnerabilities, we also found 11 devices that do not implement the DNS standards correctly: 4 routers merged CNAME chains in cached responses to a single synthesized A record, 2 routers responded with malformed packets in some cases when processing EDNS-enabled requests and 10 routers did not support connections via TCP. We also find that the inability of these routers to correctly implement some parts of the DNS standard is strongly correlated with the vulnerability to our attacks. In fact, most vulnerable devices also do not correctly implement some parts of the standard and all standard non-compliant devices are also vulnerable to one of the attacks. In terms of implementations, we find that 13 devices use dnsmasq\footnote{We identified 4 additional devices using dnsmasq in our firmware analysis in Section~\ref{sec:firmware-analysis} which do not communicate the implementation using \path{version.bind} queries.} as their DNS implementation, none of which were found vulnerable to our attacks. This means that almost all routers running a non-dnsmasq forwarder are vulnerable to one of our attacks. 3 not vulnerable devices which are not based on dnsmasq, do not implement a cache - which makes the cache-injection attacks not applicable to them.

%% file: 04-2-whitebox-firmware.tex
\section{White-Box Analysis of Routers' Firmware}\label{sec:firmware-analysis}

To find the root cause of the vulnerabilities found in Section~\ref{sec:blackbox}, we analyze the routers firmware. \new{We use firmware images downloaded from the manufacturer and source code of open source components. We then derive the conclusions on routers' behavior through a combination of white and black-box analysis to ensure accuracy.} %
We analyze the programming errors and architectural decisions which lead to the vulnerability and isolate the components which create the vulnerability. 
The results of our analysis are in Table~\ref{tab:fimrware_analysis}, along with the information about the DNS implementations in routers, information if the implementation is open source or built by the manufacturer, the vulnerabilities and incompatibilities with the standard found in the software. \new{We analyze the routers' firmware to the point where our practical results from the black box analysis can be explained by the implementation in the firmware. We also analyzed relationships between the underlying implementations and their vulnerabilities. For instance, we found some devices using the same software implementation with a different configuration which results in different vulnerability outcomes, e.g., Netgear R6120 and Tenda AC10 use dnrd.}
In our white-box analysis we add three additional routers to our dataset from Section~\ref{sec:blackbox}: Huawei 5G CPE Pro, Actiontec V1000H and Tenda AC10. These routers were tested because their firmware exhibits similarities to the routers we bought, including use of the same DNS forwarder software.

\begin{table*}[t!]
\renewcommand{\arraystretch}{0.8}
    \centering
    \footnotesize
    \begin{tabular}{r|l|c|r|l|c|c|cHH}
    \mr{2}{\textbf{Vendor}}      & \mr{2}{\textbf{Model}}              & \mr{2}{\textbf{OS}}         & \textbf{DNS forwarder} & \mr{2}{\textbf{Version}} & \textbf{Open} &  \textbf{Vulnerabilities} & \textbf{Non-} & Cache & Quirks\\ 
               &                    &            & \textbf{implementation} &    &  \textbf{source} &  & \textbf{standard} & architecture       & \\
    \hline
    \hline

 Actiontech      & MI424WR          & \mr{15}{ linux } & totd                            & 1.5            & \yes         & \vseq{} (below 1.5.3), \vport                     & \no                                          & no cache                        & \\ \cline{1-2} \cline{4-10}                                                                                                                                                           
 \mr{4}{AVM}     & Fritz!Box 7590   &                  & \mc{2}{c|}{\mr{4}{"multid"}}                 & \mr{4}{\no} & \mr{4}{\vinj}                    & \mr{4}{\no}                                          & \mr{4}{ \mytodo{(3)} }          & \mr{4}{  }  \\ \cline{2-2}                                                                                                                                                            
                 & Fritz!Box 6660   &                  &            \mc{2}{c|}{}             &              &                                  &                                           &                                 & \\ \cline{2-2}                                                                                                                                                                        
                 & Fritz!Box 7312   &                  &               \mc{2}{c|}{}            &              &                                  &                                           &                                 & \\ \cline{2-2}                                                                                                                                                                        
                 & Fritz!Box 7520   &                  &                \mc{2}{c|}{}             &              &                                  &                                           &                                 & \\ \cline{1-2} \cline{4-10} %
 Zyxel           & C3000Z           &                  & \mc{2}{c|}{\mr{2}{dproxy-nexgen}} &   \mr{2}{\yes} & \mr{2}{\vinj,\vtxid,\vport,\vcd} & \mr{2}{\vmerge,\vedns,\vntcp}             & \mr{2}{qname to address map}    & \mr{2}{\makecell[l]{CNAME merging, does not chnage txid,\\broken EDNS, UDP source port fixed after reboot}} \\ \cline{1-2} %
 Actiontec       & V1000H$^1$       &                  &                          \mc{2}{c|}{}            &              &                                  &                                           &                                 & \\ \cline{1-2} \cline{4-10}                                                                                                                                                           
 \mr{2}{Huawei}  & 5G CPE 5 Pro$^1$ &                  & \mc{2}{c|}{\mr{2}{libmsgapi.so}}        & \mr{2}{\no}  & \mr{2}{\vinj}                    & \mr{2}{\vedns}                            & \mr{2}{ \mytodo{(3)} }          & \mr{2}{\makecell[l]{DNS hijack + broken EDNS\\without internet} } \\ \cline{2-2} %
                 & 5G CPE 5 Pro 2   &                  &                           \mc{2}{c|}{}               &              &                                  &                                           &                                 & \\ \cline{1-2} \cline{4-10} %
 Cudy            & WR1300           &                  & \mr{4}{dnsmasq}                 & 2.45           & \mr{4}{\yes} & \mr{4}{\no}                      & \mr{4}{\no}                               & \mr{4}{full parsing}            & \\ \cline{1-2} \cline{5-5}                                                                                                                                                  
 
 Telekom         & Speedport Smart 3 & & & 2.59      &                               &    & &  \\ \cline{1-2} \cline{5-5}  %
 
 \mr{2}{D-Link}  & DIR-600          &                  &                                 & 2.45           &              &                                  &                                           &                                 & \\ \cline{2-2}  \cline{5-5}                                                                                                                                                           
                 & DWR-920          &                  &                                 & 2.78           &              &                                  &                                           &                                 & \\ \cline{1-2} \cline{4-10}                                                                                                                                                            
 Netgear         & R6120            &                  & \mr{2}{dnrd}                    & 2.19           & \mr{2}{\yes} & \vinj                            & \mr{2}{\no}                                          & (cache turned off in configuration)    & \\ \cline{1-2} \cline{5-5} \cline{7-7} \cline{9-10}                                                                                                                                                            
 Tenda           & AC10$^1$         &                  &                                 & 2.20.3         &              & \vinj,\vcd                       &                                           & qname to packet map             & \\ \cline{1-2} \cline{4-10}                                                                                                                                                            
 \mr{2}{TP-Link} & Archer C7        &                  & \mc{2}{c|}{\mr{2}{dnsproxy\_deamon.sh}}   & \mr{2}{\no}  & \mr{2}{\no}                      & \mr{2}{\no}                                          & \mr{2}{\makecell{no cache: forwarder\\implemented as NAT rule}}                & \mr{2}{only a NAT rule} \\ \cline{2-2}                                                                                                                                                
                 & TL-WR841N        &                  &                        \mc{2}{c|}{}              &              &                                  &                                           &                                 & \\ \hline                                                                                                                                                                             
 Strong          & Wifi Router 300  & \mr{5}{eCos}     & \mc{2}{c|}{\mr{2}{(unnamed implementation)}}    & \mr{2}{\no}  & \mr{2}{\vport}                   & \mr{2}{\vntcp}                            & \mr{2}{no cache} \\ \cline{1-2} 
 Edimax          & N300             &                  &                       \mc{2}{c|}{}            &              &                                  &                                           & \\ \cline{1-2} \cline{4-9}      
 \mr{2}{Tenda}   & F3               &                  & \mc{2}{c|}{\mr{2}{"DNS deamon"}}            & \mr{2}{\no}  & \mr{2}{\vinj,\vport,\vcd}             & \mr{2}{\vmerge,\vntcp}                    & \mr{2}{qname to address map}    & \mr{2}{CNAME merging} \\ \cline{2-2} %
                 & AC10 v3          &                  &                     \mc{2}{c|}{}              &              &                                  &                                           & \\ \cline{1-2} \cline{4-9}      
 Trendnet        & TW100-S4W1CA     &                  & \mc{2}{c|}{\mr{1}{(unnamed implementation)}}                & \no          & \vport,\vcd                      & \vmerge,\vntcp                            & \mytodo{(1)}                    & CNAME merging \\ \hline %
 Mercusys        & MW305R           & VxWorks          & \mc{2}{c|}{\mr{1}{dnsProxy.c}}            & \no          & \vseq                            & \vntcp                                    & no cache  \\ \hline             
 Bintec          & RS353a           & "BOSS"         & \mc{2}{c|}{\mr{1}{"dnsd"}}            & \no          & \vinj                            & \vntcp,\vmerge                                   & \mytodo{(3)}                    & CNAME merging, Forwards DNS messages without AD and CD Flags \\ \hline                                                                                                                
 DrayTek         & Vigor2120        & "DrayOS"       & \mc{2}{c|}{\mr{1}{(unnamed implementation)}}                       & \no          & \vinj,\vtxid,\vcd                     &  \vntcp & \mytodo{(1)}                                               & \\ \hline %
          
 Lancom          & 1640e            & "LCOS"         & \mc{2}{c|}{\mr{1}{(unnamed implementation)}}                       & \no          & -                     &  \vntcp & \mytodo{(1)}                                               & \\ \hline %

    \end{tabular} \\
    { \vinj: Misinterpretation injection. \vtxid: TXID forwarding. \vseq: Sequential TXID. \vport: Fixed UDP port. \vcd: Disable DNSSEC via CD=1. \\ \vmerge: CNAME chain merging. \vedns: EDNS can cause broken responses. \vntcp: No TCP support. $^1$: Additional router not tested physically. }
    \caption{Results of white-box analysis of routers' firmware.}
    \label{tab:fimrware_analysis}
    \vspace{-12pt}
\end{table*}

\subsection{Special character misinterpretation}
We analyze the domain name decoder and cache implementation of the routers which where found vulnerable to the special character misinterpretation with the XDRI attack. We do this to resolve the following paradox: when using a naive decoder which misinterprets special characters in domain names, this vulnerability should affect the domain name when it is received from the client in the first query and therefore trigger a query to the victim domain (\path{victim.com}) instead of the malicious attacker domain (\path{victim.com\000.attacker.com}). If implemented like this, the naive decoder would trigger a behaviour which is not compliant with the standard [RFC1035], but such a behaviour could not be exploited for an attack.

\textbf{Decoder implementation.} First, to verify that the vulnerability is indeed caused by a naive decoder logic, we analyze the domain name decoder implementations in our vulnerable routers. As expected, we find a vulnerable decoder logic in all analyzed cases: when decoding a domain name from the line format into a zero-terminated C-string, the decoders in all vulnerable routers do not handle the cases of zero-bytes nor periods inside domain name labels with special care, as a result, the name is decoded to another name than it is.

\textbf{Cache implementation.} Second, to answer why the vulnerable routers behave as they do, we analyze the cache implementations: we find 2 different types of vulnerable cache architectures, each explaining the vulnerable, non-standard behaviour of the routers we found in practice:

\textit{(a) Qname-to-address map:} Instead of parsing each record individually and moving them into a record cache, some caches in our study operate on a much simpler level. These cache implementations, which are used in both dproxy-nexgen and eCos-based Tenda devices, can be characterized as a function mapping domain names to IP addresses, and consequently, they can only store DNS records of type A (or AAAA). Queries for other types are simply forwarded to the upstream resolver every time their responses are not stored in the cache. Once a response from the upstream resolver is received, its answer section is parsed for any record of type A. Once an A record is found, it is put into the cache by mapping the domain name of the response's question section (qname) to the address found inside the (first) A record. When receiving a query from another client, the qname is decoded and looked up in the cache. If it is found, the forwarder synthesizes an A record mapping the qname directly to the cached address, otherwise it forwards the client's packet unchanged, without re-encoding the decoded name. While this represents a behaviour completely non-compliant to the DNS standard, it explains why the misinterpretation does not happen when processing the client's query and also why cached responses do not contain CNAME chains in our black-box tests.

\textit{(b) Qname-to-packet map:} Additionally, we found another cache architecture which is non-compliant to the DNS standard in dnrd. In this architecture, the cache again only considers the qname of a packet as its cache key and the forwarder does not re-encode decoded qnames when forwarding queries. However, instead of mapping qnames directly to IP addresses, this cache architecture maps qnames to full non-decoded DNS packets. When a response from an upstream resolver is received, the complete DNS packet, including its header and all sections, is cached. Then, when a query with the same qname (or a qname which is decoded to the same name because of misinterpretation) is received, the cached packet is simply sent to the client with only the TXID changed so it matches the TXID of the client's query. This causes different issues, such as flags not matching the client's query and TTLs not being reduced as the records are not updated.

\subsection{TXID forwarding}
During our analysis of caches, we also discover the main architectural cause of the TXID forwarding vulnerability. As it can already be presumed by the Qname-to-packet map cache architecture described above, many of these forwarders are not build like full-fledged resolvers, which parse all sections of a packet, interpret the parsed data, determine what has to be done to resolve a client's query and issue the respective queries to resolve the request. Instead, those forwarders, such as totd, dproxy and dnrd, are built similarly to proxies, which only extract the minimum required information from a packet which traverses it (such as the TXID and qname) and simply ignore the rest. In the cases of dproxy-nexgen and DrayTek these proxy-like forwarders do not even change the TXID itself, but simply forward the DNS packet to the upstream resolver as-is, and only tap into the response to extract the information needed for caching purposes which results in the vulnerabilities that we found.

\subsection{Missing UDP source port randomization}
During our analysis, we also identify the cause of static port numbers in queries sent to upstream resolvers. As expected, most implementations with this vulnerability, such as totd, Strong/Edimax and Trendnet implementations, simply set a static port number when creating a socket for their upstream connections, or they only create a socket once (when started) and do not recreate it for further requests, as in the case of dproxy-nexgen. However, we identified an implementation which presents an exception to this observation: the implementation in Tenda devices (F3 and AC10v3) appears to use a static port from the ephemeral port range (>=49152, see RFC6335) in our black-box test, which does not change upon reboot but is different for both devices. Port numbers in this range are typically only used by the operating systems to provide randomly assigned ports to applications which do not request a specific port themselves.

Upon analysis of the port-setup routine in the implementation which is used in both the Tenda F3 and AC10v3 routers, we find that the used port is actually chosen pseudo-randomly when the application is started, by deriving it from a call to the C \path{rand()} function. Before that, the pseudo-random generator is seeded with a call equivalent to \path{srand(time(NULL))}, which seeds the random generator with the current UNIX timestamp in seconds from January 1, 1970. This should result in a randomly chosen port every time the device is rebooted, which we do not observe in our black-box analysis. Upon further investigation, we discover, that the time on these devices is reset to January 1, 1970 after every reboot, and is only set via NTP after an Internet connection is acquired.

This explains the vulnerability we found in tested devices. Even though the UDP source port is randomly chosen once the device is started, at this point the clock in the router is still set to some seconds after January 1, 1970. The different ports chosen for the different routers (F3: 50387, AC10v3: 62066) are due to the AC10v3 starting its DNS forwarder slightly later after the clock has been reset: by extracting the code deriving the port from the current time, we run it with values for \path{time()} between 0 and 100 and find that it generates the port 50387 with a value for \path{time()=7} and 62066 with \path{time()=10}, concluding that the DNS forwarder in the Tenda AC10v3 is started 10 seconds after the device has been started, vs. 7 seconds in the case of the Tenda F3. As these devices use an embedded operating system (eCos), these boot times are deterministic enough so that the port number generated does not change between reboots.

To summarize, there are 3 vulnerabilities/bugs, alone in the code described above: (1) not changing the UDP source port for different requests, (2) making a security-relevant random number (UDP source port) only dependent on the current time (in seconds) and (3) not acknowledging that the device's clock is not correctly set up at the time when the port is chosen, making the pseudo-random generator behave completely deterministic. These examples are representative of our implementations we reverse engineered.

\subsection{CD=1 forwarding}
\new{The goal of the white-box analysis of the CD flag is to understand why a given vulnerability exists. The results of the white-box analysis allow us to develop the attacks in Section \ref{sec:attacks}.} Similar to the TXID forwarding vulnerability, the main cause of checking disabled forwarding in vulnerable routers is due to the fact that these routers do not recognize the CD flag in the DNS flags field. In our analysis, we found multiple ways how forwarders handle this situation. The forwarders in Bintec clear the CD flag (along with other unrecognized flags) when forwarding queries. For resolvers not supporting this flag, it is a standard-compliant solution [RFC1035, RFC8906], but prevents clients from issuing queries with these flags set.

Instead of clearing unknown flags, the Trendnet, DrayTek, Tenda and dproxy-nexgen forwarders do not clear unrecognized flags but simply forward them without interpreting their meaning. This leads to the vulnerability as the responses for queries with these flags are processed normally and the invalidated records are stored in the forwarder's cache.

Finally, instead of clearing the flag, the forwarders in AVM, Huawei and dnsmasq handle the CD flag by forwarding queries with it transparently. However, instead of parsing their responses and potentially poison their cache with DNSSEC-signed records that were not validated, they only forward responses with the flag set to the client requesting them, but do not copy the included records into their cache. This can be seen as a way of providing some forward-compatibility for their clients without actually implementing the required functionality.

\subsection{Missing TCP support}

Even though a mandatory requirement in [RFC5966], many routers do not support TCP queries. Our firmware analysis shows, that while many implementations indeed do not support DNS transport over TCP, such as dproxy, Edimax/Strong, Mercusys and Tenda, in some cases TCP is not enabled due to bugs. For instance, in the case of totd, used in the Actiontech MI424WR, the application actually does support TCP transport, however it is deactivated in the configuration of the DNS forwarder as it is used in the router firmware.

\subsection{Analysis of DNS software on github}\label{ref:github-study}

\begin{table}[t!]
\renewcommand{\arraystretch}{0.8}
    \centering
    \footnotesize
    \setlength{\tabcolsep}{5.5pt} %
    \begin{tabular}{Hl|c|c|cccc|c|c|cH}

\multirow{4}{*}{ \makecell[c]{Programming\\language} } &                              &       & \multicolumn{8}{c}{Attacks} & \\ \cline{4-11}
         & \multirow{3}{*}{ \rot{\makecell[c]{Project ~ }} }    & \multirow{3}{*}{ \rot{Cache ~ } }    & \multirow{3}{*}{ \rot{Any ~ ~} }    & \multicolumn{4}{c|}{Misinterpretation} &  \multirow{3}{*}{ \rot{\makecell[c]{TXID\\forwarding}} } & \multirow{3}{*}{ \rot{Fixed port} } & \multirow{3}{*}{ \rot{\makecell[c]{Disable\\DNSSEC\\via CD=1}} } & \\
         &                                    &     &      & \multicolumn{2}{c}{direct} & \multicolumn{2}{c|}{CNAME} &  &  &   & \\
         &              &   &    & \rot{\path{\.}} & \rot{\path{\000}} & \rot{\path{.}} & \rot{\path{\000}} &     &     &   & notes \\ 
\hline
C        & \cite{aa65535/hev-dns-forwarder}          & \no    & \yes    & \no$^4$ & \no$^4$& \no$^4$ & \no$^4$& (\yes)$^4$ & seq.  & \no$^4$     &          \\ %
go       & \cite{Yangshifu1024/dnsforwarder}         & \yes   & \yes    & \no     & \no    & \no     & \no    & \yes    & \no      & \yes     &  \\
go       & \cite{Sina-Ghaderi/nanodns}               & \yes   & \no     & \no$^7$ & \no$^7$ & \no     & \no    & \no     & \no      & \no$^5$ & CNAME merging        \\
go       & \cite{sodapanda/doublebarrel}             & \yes   & \yes    & \no     & \no    & \no     & \no    & n/a$^3$ & n/a$^3$  & \yes     &  \\
nodejs   & \cite{yyfrankyy/fwdns}                    & \yes   & \yes    & \no$^1$ & \no    & \no$^2$ & \no    & \no     & \yes$^6$ & \no$^5$  &  \\  %
go       & \cite{wolf-joe/ts-dns}                    & \yes   & \yes    & \no     & \no    & \no     & \no    & \yes    & \no      & \yes     &             \\
python   & \cite{xiaoyang-sde/encrypted-dns}         & \yes   & \yes    & \no     & \no    & \no     & \no    & \yes    & \no      & \yes     &  \\
go       & \cite{janeczku/go-dnsmasq}                & \yes   & \yes    & \no     & \no    & \no     & \no    & \yes    & \no      & \yes     &      \\  %
ruby     & \cite{arloan/prdns}                       & \no    & \no     & \no$^4$ & \no$^4$& \no$^4$ & \no$^4$& \no$^4$ & \no      & \no$^{4,5}$  &  \\ %
go       & \cite{import-yuefeng/smartDNS}            & \yes   & \yes    & \no     & \no    & \no     & \no    & \yes    & \no      & \yes     &  \\     %
\hline
    \end{tabular} \\
    {\yes: vulnerable/yes. \no: not vulnerable/no. $^1$: misinterpreted but not cached. \\ $^2$: misinterpreted before forwarding. $^3$: off-path attack prevented with DoT. \\ $^4$: attack (also) prevented by not implementing a cache. $^5$: CD flag stripped. $^6$: port only selected randomly once. $^7$: no answer. }
    \caption{Vulnerabilities in github DNS forwarder software.}
    \label{tab:github_forwarders}
    \vspace{-12pt}
\end{table}

During our study, we found a significant number of vulnerable DNS forwarder implementations in routers, with both open-source and closed-source implementations being affected. %
To understand why errors and bugs are so frequent in DNS implementations we additionally \new{perform black-box evaluations} of self-built DNS forwarders by independent github developers. We search github for "dns forwarder" keywords and select 10 small-size projects which implement DNS forwarders. We test these implementations for the vulnerabilities listed in Section~\ref{sec:attacks}. The results are summarized in Table~\ref{tab:github_forwarders}. Out of 10 tested implementations, only 2 are resilient to our attacks: one simply does not implement a cache, and the other one still does not implement the DNS standard correctly as it merges CNAME chains in responses to a single A record.

While these implementations are not widely used, the fact that almost all of them carry similar vulnerabilities to the implementations used by the vulnerable routers in our dataset shows that DNS forwarding is not easy to implement and that the programming mistakes leading to the attacks are common and not just individual cases. %

%% file: 05-adnet-study.tex
\section{Advertisement Network Study}
\label{sec:adnet-study}

\begin{figure}[t]
    \centering
    \includegraphics[width=0.43\textwidth]{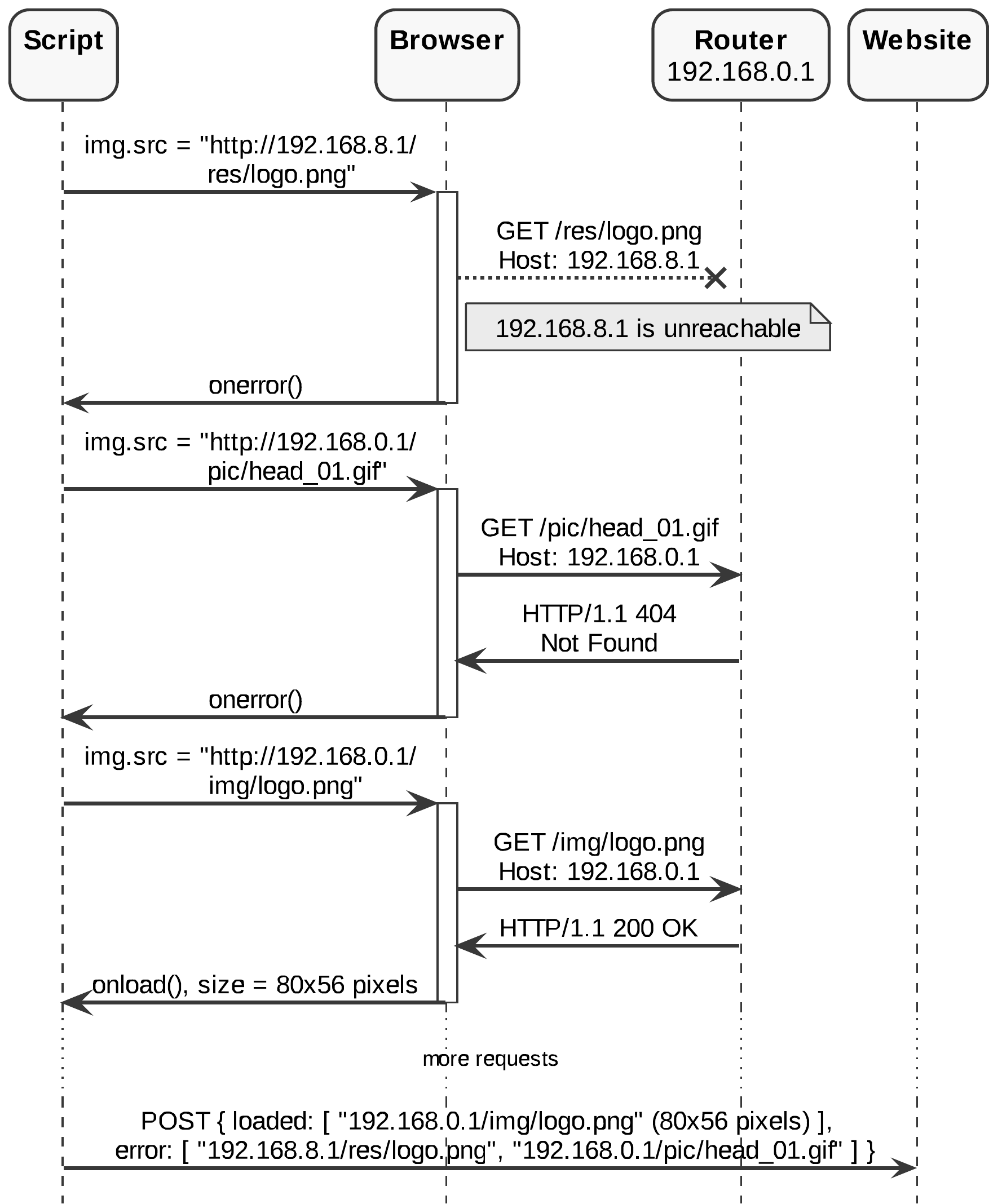}
    \caption{\new{Fingerprinting routers from javascript.}}
    \vspace{-10pt}
    \label{fig:adnet_study}
\end{figure}

We develop novel techniques to fingerprint residential routers, which we then apply in our Internet wide ad network study. We embed a javascript on our website which is loaded as a pop-under advertisement and executes our techniques for identifying routers. \new{This technique is shown in Figure~\ref{fig:adnet_study}.}

We analyze the web interfaces of the routers in our dataset and create a list of images specific to each router. Once a client has loaded our website, the script tries to fetch an image from characteristic URLs from \new{each of} the routers' web-interfaces \new{by setting the \path{src} property of a newly created image object}. Since these images only exist on the web interface specific to one router, the ability to successfully load such an image implies that the client is using the router in question. \new{During the execution the script loads all the images and reports back for which images the tests succeeded. Same models from the same vendor might use the same images, but in this case they typically use the same firmware and the same vulnerable DNS implementation.}

To find the router's address, we use the default address shipped with the factory settings. Additionally we use the special domain names used by some vendors to redirect users to the router's web interface. DNS requests for such domains, like \path{"routerlogin.net"} or \path{"fritz.box"} are always answered with a router's internal IP by the router's DNS forwarder. 

\new{We confirm experimentally on all the routers in our study, that firmware upgrades do not change the images of web configuration interfaces, and hence do not have an impact on our fingerprinting heuristics. In first round of tests we test the routers in the default factory settings configurations, as this is likely how they are used by the clients. In a subsequent round we update the routers, but do not perceive any changes in the results. In addition, not all the tested routers allow the user to manually update the firmware. Notice that if such changes existed our heuristic would result in errors. For instance, if the ports were changed but the images and the domains remained the same, this would create false positives, alternately, if the image and domains are changed but ports remain static, this would create false negatives.}

To circumvent the restrictions of the same-origin-policy (SOP), our script is only loading the characteristic URLs by creating image objects, which are exempt from the SOP. While this circumvents the restrictions of the SOP, this limits the information which is gathered by our script to a few limited data points: whether an image was successfully loaded or not and the dimensions of the image in pixels. Therefore, to increase the accuracy of our study, once an image is loaded successfully, we compare its dimensions to the ones we found using our initial analysis of the routers' web-interfaces.

To find out image paths and characteristic for each router, we manually search the web interface of each router for images embedded in the webpage which are specific for that router. During this examination, we find that some routers prevent images from being loaded by third-party websites, either directly or indirectly, by enforcing that the HTTP requests referrer header must match the IP address of the router\footnote{Some routers did enforce a weaker policy where they also allow empty referrer headers. This is not enough to block our study since sending of this header can be disabled by setting the \path{referrerPolicy} attribute of an HTML image to \path{no-referrer}.}, enforcing a connection via HTTPS using a self-signed certificate or always requiring authentication even for static objects of the web interface. In total, we were able to gather fingerprints in the form of characteristic image paths for 27 out of the 36 routers in our dataset.

\begin{table}[t!]
\renewcommand{\arraystretch}{0.6}
\footnotesize
    \begin{newenv}
    \centering
    \footnotesize
    \setlength{\tabcolsep}{3.5pt}
    \begin{tabular}{r|rr|rr|r}
         & \multicolumn{2}{c|}{\textbf{All clients}} & \multicolumn{2}{c|}{\textbf{Identified}}         & \\
         & \textbf{absolute}	 &\textbf{\% of total} & \textbf{absolute} &\textbf{\% of total} & \textbf{\% Identified} \\
\hline
Europe   & 9608	     & 14.04\%    & 176      & 18.09\%    & 1.83\% \\
Asia     & 24516	 & 35.81\%    & 348      & 35.77\%    & 1.42\% \\
Americas & 11773	 & 17.20\%    & 181      & 18.60\%    & 1.54\% \\
Africa   & 15669	 & 22.89\%    & 230      & 23.64\%    & 1.47\% \\
Oceania  & 1225	     & 1.79\%     & 19       & 1.95\%     & 1.55\% \\
Unknown	 & 5664	     & 8.27\%     & 23       & 2.36\%     & 0.41\% \\
\hline
Total	 & 68455	 & 100.00\%  & 973       & 100.00\%   & 1.42\% \\
    \end{tabular}

    \caption{Adnet results by region}
    \label{tab:adnet_regions}
    \end{newenv}
\end{table}

\textbf{Results.}
\new{In total, 68,455 clients from 202 different countries and 4653 different autonomous systems visited our website during the study. Out of those, we could identify a known router web interface for 973 (1.42\%) clients, which are located in 438 different autonomous systems and 105 countries. A further geographical breakdown of our results is shown in Table~\ref{tab:adnet_regions}.}

The number of matches for each web interface is shown in Table~\ref{tab:adnet_routers}. When we count the number of instances in which we found vulnerabilities in the router, we find that in 929 (or 1.36\%) of the total cases, a vulnerable router was identified. This is partly since these vendors use the same web-interface for all their router models, which increases the number of identified devices for these vendors. As can be seen in Table~\ref{tab:adnet_routers} the more generic the match is the higher \% of routers can be identified. \new{“Generic match” means the match applies to different models of the same vendor, which may or may not be vulnerable as well.}

\begin{table}[t!]
\renewcommand{\arraystretch}{0.8}
    \centering
    \footnotesize
    \setlength{\tabcolsep}{3.5pt}
    \begin{tabular}{r|r|r|c|c|l}
\textbf{\% of}	   & \textbf{\% of}	&     	& \textbf{Generic} 	& \textbf{Vulne-}	&  \\ 
\textbf{Identified} & \textbf{Total}	& \textbf{Absolute}	& \textbf{match}	& \textbf{rable}	& \textbf{Router} \\ 
\hline
41.62\%	 & 0.59\%	 & 405	 & x	 & x	 & Tenda \\ 
24.25\%	 & 0.34\%	 & 236	 & x	 & x	 & Huawei \\ 
15.42\%	 & 0.22\%	 & 150	 & x	 & x	 & Fritzbox \\ 
12.13\%	 & 0.17\%	 & 118	 & x     & x	 & Mercusys \\ 
1.54\%	 & 0.02\%	 & 15	 & x	 & 	     & Linksys AC2200 \\ 
1.34\%	 & 0.02\%	 & 13	 & 	     & 	     & Xiaomi Mi router \\ 
1.23\%	 & 0.02\%	 & 12	 & 	     & x	 & Draytek \\ 
1.13\%	 & 0.02\%	 & 11	 & 	     & 	     & Speedport Smart \\ 
0.72\%	 & 0.01\%	 & 7	 & 	     & x	 & Netgear R6120 \\ 
0.21\%	 & 0.00\%	 & 2	 & 	     & 	     & D-Link DIR-600 \\ 
0.10\%	 & 0.00\%	 & 1	 & 	     & 	     & Teltonika \\ 
0.10\%	 & 0.00\%	 & 1	 & 	     & 	     & Linksys AC1000 \\ 
0.10\%	 & 0.00\%	 & 1	 & 	     & x	 & Centurylink \\ 
0.10\%	 & 0.00\%	 & 1	 & 	     & 	     & ASUS ROG \\ 
\hline
100.00\%	 & 1.42\%	 & 973	 & 	- & 95.48\%	 & Identified \\ 
	- & 98.58\%	 & 67482	 & 	- & -	 & Not Identified \\ 
\hline
	- & 100.00\%	 & 68455	 & 	- & 1.36\%	 & Total %
    \end{tabular}
    \caption{Identified routers from ad-net clients}
    \label{tab:adnet_routers}
\end{table}

\begin{table}[t!]
\renewcommand{\arraystretch}{0.8}
    \centering
    \footnotesize
    \setlength{\tabcolsep}{3.5pt}
    \begin{tabular}{r|r|r|l}
\textbf{\% Identified}	 & \textbf{Frequency}	 & \textbf{\% of Total}	& \textbf{Browser} \\ 
\hline
0.43\%	 & 22618	 & 33.04\%	 & Chrome\\ 
3.82\%	 & 13633	 & 19.92\%	 & Chrome Mobile WebView \\ 
0.80\%	 & 8764	     & 12.80\%	 & Chrome Mobile\\ 
1.18\%	 & 4569	     & 6.67\%	 & Mobile Safari \\ 
1.63\%	 & 3076	     & 4.49\%	 & Safari\\ 
0.79\%	 & 3033	     & 4.43\%	 & Edge\\ 
1.56\%	 & 2824	     & 4.13\%	 & Firefox\\ 
1.98\%	 & 2270	     & 3.32\%	 & Samsung Internet \\ 
1.94\%	 & 1597	     & 2.33\%	 & Firefox Mobile \\ 
\hline
1.42\%	 & 68455	 & 100.00\%	 & Total%
    \end{tabular}
    \caption{Browsers of ad-net clients}
    \label{tab:adnet_browsers}
    \vspace{-10pt}
\end{table}

We also list the distribution of clients browsers in Table~\ref{tab:adnet_browsers}. The number of cases where we could identify a router web interface is below the average of 1.42\% in the case of Chrome (both desktop and mobile) and Edge, as both of these browsers block the requests to internal IP addresses \cite{chrome-privatenetworkaccess,w3c-privatenetworkaccess}. The fraction of devices we could identify with our study is significantly lower than the fraction of routers vulnerable to our attacks. In a real attack the limitations to fingerprinting do not reduce the attack success.

The distribution of the vulnerable devices according to countries and ASes from the ad network study compared to routers with open DNS forwarders in Table \ref{tab:scan_table}. The overlap between the ASes with vulnerable devices measured with ad net vs devices with open resolvers \new{as well as the total number of scanned ASes} is shown in Table~\ref{tab:vuln_asns_open_adnet}.

\begin{table}[t!]
\renewcommand{\arraystretch}{0.8}
    \centering
    \footnotesize
    \setlength{\tabcolsep}{3.3pt}
\begin{tabular}{l|r|r|r|r|r|r}
\textbf{(1) Open res.}&\mc{2}{c|}{\textbf{Ad-net}}&\mc{2}{c|}{\textbf{Open resolver}}&\mc{2}{c}{\textbf{Vulnerable of Total (\%)}}\\
\hline

&&\textbf{Vulne-}&&\textbf{Vulne-}&&$\downarrow$ \textbf{Open}\\

\textbf{Country}&\textbf{Total}&\textbf{rable}&\textbf{Total}&\textbf{rable}&\textbf{Ad-net}&\textbf{ resolver}\\

\hline
\hline
Oman&75&2&1295&1230&2.67&94.98\\
\hline
\makecell[l]{Papua New\\ Guinea}&33&1&55&22&3.03&40.00\\
\hline
Rwanda&50&0&80&29&0.00&36.25\\
\hline
Iceland&1&0&144&50&0.00&34.72\\
\hline
Nicaragua&19&0&470&159&0.00&33.83\\
\hline
Bahrain&26&1&134&43&3.85&32.09\\
\hline
Thailand&3410&14&13011&3716&0.41&28.56\\
\hline
Guadeloupe&15&0&73&20&0.00&27.40\\
\hline
Macao&0&0&374&102&0.00&27.27\\
\hline
Singapore&45&0&5560&1498&0.00&26.94\\
\hline
Croatia&10&0&780&206&0.00&26.41\\
\hline
Luxembourg&4&0&214&55&0.00&25.70\\
\hline
\makecell[l]{
United Arab\\Emirates}&321&3&1628&407&0.93&25.00\\
\hline
Guernsey&0&0&57&14&0.00&24.56\\
\hline
Japan&224&0&19151&4543&0.00&23.72\\

\mc{7}{c}{~} \\
\mc{7}{c}{~} \\

\textbf{(2) Ad-net}&\mc{2}{c|}{\textbf{Ad-net}}&\mc{2}{c|}{\textbf{Open resolver}}&\mc{2}{c}{\textbf{Vulnerable of Total (\%)}}\\
\hline

&&\textbf{Vulne-}&&\textbf{Vulne-}&&\textbf{Open}\\

\textbf{Country}&\textbf{Total}&\textbf{rable}&\textbf{Total}&\textbf{rable}& \hspace{5pt} $\downarrow$ \textbf{Ad-net}& \textbf{resolver}\\

\hline
\hline
Kuwait&58&17&914&40&29.31&4.38\\
\hline
Germany&285&78&13627&939&27.37&6.89\\
\hline
Namibia&75&19&162&8&25.33&4.94\\
\hline
Botswana&63&6&365&10&9.52&2.74\\
\hline
Bangladesh&456&36&17524&112&7.89&0.64\\
\hline
Honduras&65&5&1687&14&7.69&0.83\\
\hline
Iraq&487&33&1283&27&6.78&2.10\\
\hline
Panama&74&5&1801&32&6.76&1.78\\
\hline
Qatar&60&4&103&12&6.67&11.65\\
\hline
New Zealand&139&9&1874&292&6.47&15.58\\
\hline
South Africa&1719&110&8785&1139&6.40&12.97\\
\hline
Saudi Arabia&368&22&2021&118&5.98&5.84\\
\hline
Ecuador&172&9&3422&180&5.23&5.26\\
\hline
Pakistan&518&25&4562&105&4.83&2.30\\
\hline
\makecell[l]{
Trinidad\\and Tobago}&448&19&205&7&4.24&3.41\\

\end{tabular}
    \caption{Vulnerable open resolvers and routers by country, sorted by \% of vulnerable (1) open resolvers and (2) ad-net routers.}
    \label{tab:scan_table}
    \vspace{-10pt}
\end{table}

\begin{table}[t!]
\renewcommand{\arraystretch}{0.9}
    \centering
    \footnotesize
    \setlength{\tabcolsep}{1.9pt}
    \begin{tabular}{r|c|c|c|c}
    & \textbf{only open}      & \textbf{open resolver} & \textbf{only} & \multirow{2}{*}{\textbf{total}} \\
    & \textbf{resolver} & \textbf{$\cap$ ad-net}     & \textbf{ad-net} & \\
    \hline
    \new{total scanned} & \new{18812 ASes}     & \new{3591 ASes}           & \new{1062 ASes} & \new{23465 ASes} \\
    \hline
    contains vulnerable & 4418 ASes     & 206 ASes           & 205 ASes & \new{4829 ASes} \\
    resolvers or routers   & \new{23.5\%}            & \new{5.7\%}                 & \new{19.3\%} & \new{20.6\%} \\
    \end{tabular}
    \caption{ASes in open resolver and ad-net studies}
    \label{tab:vuln_asns_open_adnet}
    \vspace{-12pt}
\end{table}

%% file: 06-nodns.tex
\section{Removing DNS From Routers} \label{sc:unorthodox}\label{sec:routerwithoutdns}
In this section we provide recommendations for mitigating our attacks. We also describe a more systematic solution for enhancing resilience of the routers. 

{\bf Recommendations for mitigations.} We set up a tool to enable routers' developers as well as anyone else to test the residential routers on their networks for vulnerabilities that expose to our attacks at {\small{\path{https://xdi-attack.net/}}}. To avoid vulnerabilities we recommend that developers of routers use well known DNS implementations, such as dnsmasq or bind, and not implementations that were not extensively analyzed, like those that we found on github (listed in Table \ref{tab:github_forwarders}). %
We provide more details and additional recommendations for countermeasures in Appendix, Section \ref{sc:mitigations}. Nevertheless, the different vulnerabilities and bugs found in our as well as previous works show that DNS is a weak spot in routers. Removing DNS improves the resilience of routers and client networks, without a significant loss of performance. We next explain the role of DNS in routers and propose simple solutions for removing them.

{\bf Role of DNS in residential routers.} Our study shows that many residential routers do not implement most of the functionalities and security features of DNS, yet they still contain DNS forwarders which merely proxy the clients' requests to another address. These forwarders only separate the network configuration of the internal client network from the network of the ISP. This provides connectivity without the need for synchronization. An interested reader is referred to Section \ref{ap.sc.examples} in Appendix, for example problems that can occur due to lack of synchronization. We show that the synchronization issue can be easily solved without a DNS in router. \new{Our recommendation to remove DNS applies only to DNS components in routers as our study has shown these are the types of devices most affected by our attacks. For other devices, such as ISP resolvers, implementations should be patched to remove the vulnerability.}

{\bf How to eliminate DNS from routers.} We provide two approaches for removing DNS. One is a simple configuration: on the client side the users should configure their systems to send queries to an upstream resolver directly, such as the ISP's resolver. The clients can then bypass the DNS forwarders in routers. The other approach offers a more systematic method for implementing this, without requiring the clients to change their network configurations. The idea is that instead of implementing a DNS forwarder as a service running on the router's operating system, the routers can implement DNS forwarding as a simple Network Address Translator (NAT) rule. The clients behind the routers are still configured via DHCP to send their DNS requests to the router's internal network address, but these packets are sent directly to the ISP's resolver by rewriting their destination addresses (and potentially also the source ports by the NAT) with a firewall rule. Each time the Internet connection is re-established, e.g., when the router is booted, the firewall rule is changed to route the DNS requests to the correct destination. In fact, in our white-box analysis we found one implementation (TP-Link) to be already doing this instead of implementing an internal DNS forwarder.

\begin{figure}[t!]
    \centering
    \includegraphics[width=0.42\textwidth]{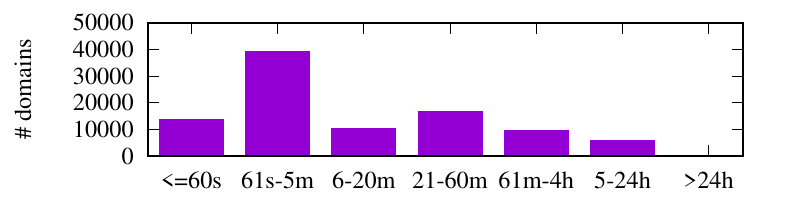}
    \caption{TTL of Alexa top 100K domains' A records.}%
    \label{fig:alexa100k_ttl}
    \vspace{-12pt}
\end{figure}

{\bf Performance costs of eliminating DNS from routers.} The routers offer caching to the client devices on a local network. This reduces the latency to the resolver of the ISP. However, the resolvers of the ISPs are located in proximity to client networks and hence latency is limited to milliseconds. Additionally, the resolver of the ISP has a much larger cache, which also includes the records cached on the residential router. Hence, \begin{newenv}to estimate the performance hit when removing DNS caching from routers we consider 2 scenarios:

\textit{a) Full-fledged operating systems, which include their own local DNS cache.} In this case the only performance hit is an increase in page loading times of 1 RTT, measured between the local device and the ISP's resolver, when a DNS record is requested for which there is no cached corresponding record in the local DNS cache, but is present on the router's cache. This is only the case when the record in question was previously requested by another device on the local network and as long as the record has not timed out yet.\end{newenv}
Since the TTL values of most records are less than 5 minutes, in most cases even with DNS cache in router, the clients still needs to wait for the query to be recursively resolved by the ISP. We measured the TTL values of popular A records in Alexa 100K-top domains, and found that 14.066 domains (14\%) have a TTL of up to 60 seconds and 53473 (55\%) have a TTL under 5 minutes, see Figure~\ref{fig:alexa100k_ttl}. \begin{newenv}
Consequently, for a router's DNS cache to save the loading time of a typical web-page, the user has to load a page which he has not visited for the last 5 minutes (otherwise the record is locally cached), which was also visited by another device on the same local network less than 5 minutes ago, so that the record is present in the router's cache and has not timed out yet. %

\textit{b) Small, IoT-like devices without a local DNS cache.} In this case, as the device does not include a local DNS cache, the removal of the DNS cache from the router means that every DNS query is sent to the ISP's resolver. This results in a small increase in connection establishment latency for those devices and a potentially higher load on the ISP's resolver. The increased latency in the range of 1 RTT will not affect most IoT applications as there is no user involved to perceive the increased interaction latency, but the increased load on ISP's resolvers might result in slightly higher operation costs at the ISP.
\end{newenv}

\new{{\bf Security benefits.} Removing DNS from routers solves not only the injection attacks, but also prevents entropy reduction via port or TXID prediction, since there is no shared cache anymore which reuses the port or TXID selected by the attacker. In addition, the DNSSEC CD flag can also not be turned off, since it will not be set by the router.} %

%% file: 07-conclusions.tex
\section{Conclusions}\label{sc:conclusions}
We show that popular routers, deployed in the Internet, are vulnerable to DNS cache poisoning attacks. We found the vulnerabilities in multiple countries and networks, which shows that these are not individual cases and the problems are prevalent. We perform the first systematic analysis of the factors causing the vulnerabilities and find that the problems are, among others, due to incorrect assumptions about the networks, developers errors, misunderstanding of the standard, simplified DNS implementations. These issues are present in different open and closed source implementations we analyzed. The high competition over the markets causes the routers' vendors to "cut corners". %
However, DNS resolvers are complex systems, with many components, simplification or modification of any of these can lead to vulnerabilities, which is indeed what our evaluations find. 

We provide recommendations for countermeasures and suggest a change of paradigm for making networks and routers more resilient: to remove DNS from the routers.

%% file: 09-appendix.tex
\section{Cache Poisoning via BGP Hijacking}\label{ap.sc.cache.poison}

In this section we provide details on how an attacker can conduct a cache-poisoning attack against a router which has the CD forwarding vulnerability by conducting an off-path attack against the upstream resolver even when this resolver it performing DNSSEC validation and the victim domain is DNSSEC-protected.
The attack is shown in Figure~\ref{fig:attack_cd} where the attacker launches the attack against the upstream resolver in Step 4.

In practice such attacks are often launched via interception of DNS request with BGP prefix hijacks \cite{myetherwallet,dns:venezuela}. Normally the DNSSEC validation would fail and the record injected by the adversary would be discarded, but in this case it is forwarded to the router (without it being cached by the upstream resolver) because it has the CD flag set. The vulnerable router receives the injected packet and caches the address \path{6.6.6.6} for the domain \path{victim.com}, because it does not know about the meaning of the CD flag and simply ignores it.

When the client visits the website at \path{victim.com}, this triggers a query for \path{victim.com} without the CD flag set. Because the router did not check the CD flag previously and put the unchecked record in it's cache, it responds with the injected record, which sends the client to the attackers server at \path{6.6.6.6}.

%% file: 05-5-countermeasures.tex
\section{Recommendations for Patches}\label{sc:mitigations}

In this section, we discuss countermeasures against the vulnerabilities found in this study which can be implemented by different parties. We first discuss countermeasures which can be implemented by users and third-party resolver operators and then present ways to reduce the likeliness of such vulnerabilities which can be implemented by router manufacturers.

\textbf{End-users.} The simplest recommendation for end-users is to not use the DNS forwarder in their router, but configure their systems to send their queries to a non-vulnerable resolver directly, such as the ISPs resolver, or a public open resolver. To test if their router is vulnerable to the attacks, we provide a test application which can detect the misinterpretation injection and CD forwarding vulnerabilities\footnote{Unfortunately, the TXID forwarding and static UDP port vulnerabilities cannot detected without access to the forwarders upstream connection.}. %

\textbf{Upstream resolver operators.} Operators of upstream resolvers can protect their users against some of the attacks by not forwarding the malicious injection queries required to conduct the misinterpretation attack and to disrespect the Checking Disabled flag. However, both of these option represent violations of the DNS or DNSSEC standards and could cause collateral damage in the form of no being able to serve other clients correctly anymore.

\textbf{Forwarder developers.} Developers of the vulnerable forwarding software should fix the vulnerabilities which allow these attacks. However, security-concerned router manufacturers might want to consider replacing their self-built DNS forwarder implementations with well-known implementations such as dnsmasq or bind. While there are also vulnerabilities in these well known DNS resolver implementations as well, these implementations are reviewed for vulnerabilities by third parties and vulnerabilities are typically fixed in a timely manner once they are found. On the other hand, smaller self-build implementations have a higher likelihood of containing programming errors and vulnerabilities in them will often not be discovered for a long time: For example, in Section~\ref{ref:github-study} we analyze self-build DNS forwarder implementation on github and find that almost all of them contain at least one of the vulnerabilities from Section~\ref{sec:attacks}.

However, shipping security updates for updated DNS resolver implementations regulary might present a challenge to router manufacturers, as fixed implementations must be tested and firmware updates have to be build and installed on the affected devices. For example, out of the devices running dnsmasq in our study, the implementation in 7 out of 17 devices is older than 7 years when running the out-of-box firmware.
As an alternative for using well-known forwarder implementations, router manufacturers might want consider removing the DNS cache from the router altogether in an approach to reduce attack surface like described in Section~\ref{sec:routerwithoutdns}.

\textbf{Stub-resolvers.} As all of the attacks evaluated in this work aim to interfere with the DNS transaction data integrity, DNSSEC is theoretically able to detect such changes if validated by vulnerable routers' clients. However, in the currently prevalent DNSSEC deployment model, DNSSEC validation is only performed by the recursive resolver ran by ISPs or bigger network operators, so changes to the DNS records which occur after the data leaves these resolver cannot be detected. A change of this paradigm, where validation is conducted on the end systems, i.e., the stub resolvers on the client itself would therefore prevent these attacks. There are technical challenges for DNSSEC validation in stub resolvers as not all recursive resolvers or forwarders forward the required signatures to the end systems to be able to perform validation [RFC8027]. As we found, this is also true for many of the vulnerable implementation in the routers we tested as these forwarders are not DNSSEC aware and remove record types they assume are not required to resolve a query.

%% file: synchronization.tex
\section{DNS in Routers: What and Why}\label{ap.sc.examples}

In a typical DNS forwarder setup on a residential router, the router creates its own internal network segment (\path{192.168.0.0/24}) from which internal clients are allocated IP addresses, we provide an illustration in Figure~\ref{fig:dnsforwarding}. The router configures all the clients on the network to send their DNS requests to the router's address (\path{192.168.0.1}) via DHCP. To handle those requests, instead of implementing a recursive DNS resolver in the router itself, the router forwards these requests to the upstream DNS resolver of the Internet service provider (ISP) at \path{198.51.100.34}. This address the router has received from the ISP when setting up the Internet connection via PPPoE or DHCP. This resolver operates in a recursive mode and looks up the appropriate authoritative nameservers for each query via DNS and sends the queries to them, e.g., the authoritative nameserver for \path{example.com} at \path{203.0.113.45}.

\begin{figure}[t!]
    \centering
    \includegraphics[width=0.47\textwidth]{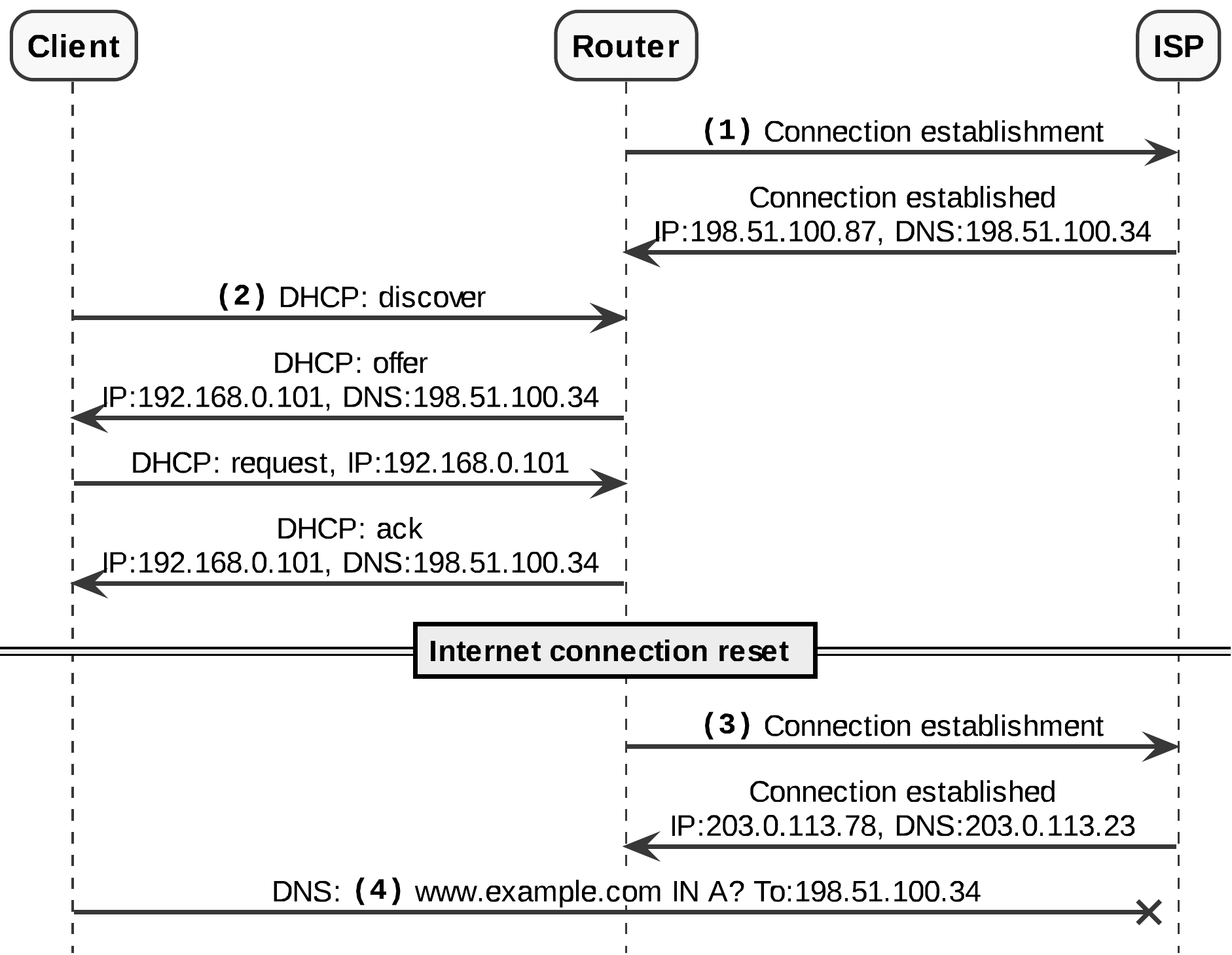}
    \caption{Why routers contain DNS forwarders.}
    \label{fig:why-dns-dhcp}
\end{figure}

{\bf Why DNS forwarders in routers are needed?} To show why DNS forwarding is needed, we illustrate the following scenario in Figure~\ref{fig:why-dns-dhcp}. In this scenario, we show why routers do need to implement some form of DNS forwarding and cannot simply hand over the address of the ISPs DNS resolver to their clients via DHCP.

In the first step (1), the router establishes an Internet connection to the ISP (i.e., via PPPoE) and receives the address configuration from the ISP, containing the instruction to set the DNS server to \path{198.51.100.34}.

In step (2), a client connects to the routers network and requests its address configuration via DHCP. The router provides the configurations and tells the client to send its DNS requests to the ISP DNS server directly at \path{198.51.100.34}.

Afterwards, the internet connection is reset and reestablished in step (3). However, the ISP has switched over the customer to a different network block, this time configuring the router to send it's DNS requests to another DNS resolver at \path{203.0.113.23}. This situation can also happen when the router switches an internet connection (some routers offer dual-internet connection support) or when the internet connection is not available yet when a client connects (In this case the DNS configuration in the DHCP transaction would be empty).

Now, in step (4), the client which did not notice the Internet reconnect issues a DNS request to its configured DNS server at  \path{198.51.100.34}, however this server is not reachable anymore because of the different ISP configuration. As DHCP does not offer any means to communicate the changed address of the DNS server to the client, the only option to resolve the situation is to reconnect the client to the network.

